\documentclass[10pt,prd,twocolumn,nofootinbib,notitlepage,aps,tightenlines,floatfix,
preprintnumbers,amsmath,amssymb,amsfonts,showpacs,superscriptaddress]{revtex4-1}

\RequirePackage[colorlinks=true
,urlcolor=blue
,anchorcolor=blue
,citecolor=blue
,filecolor=blue
,linkcolor=blue
,menucolor=blue
,linktocpage=true
,pdfproducer=medialab
,pdfa=true
]{hyperref}

\usepackage{graphicx,epstopdf}
\usepackage{mathtools}
\usepackage{comment}
\usepackage{natbib,bm}
\usepackage[usenames,dvipsnames]{color}
\usepackage{slashed}    
\usepackage{xcolor}
\usepackage{multirow}
\usepackage{grffile}
\usepackage{physics}
\usepackage{dsfont}
\usepackage[]{hyperref}
\hypersetup{
	colorlinks,
	linkcolor={blue},
	linktocpage=true,
	citecolor={blue},
	urlcolor={blue}
}
\usepackage[capitalize]{cleveref}

\usepackage{mathtools, nccmath}
\usepackage{lipsum}
\usepackage{cancel}
\usepackage[normalem]{ulem}
\usepackage{chngcntr}

\def\be{\begin{equation}}
\def\ee{\end{equation}}
\def\ba{\begin{eqnarray}}
\def\ea{\end{eqnarray}}
\def\ge{\mathrel{\raise.3ex\hbox{$>$\kern-.75em\lower1ex\hbox{$\sim$}}}}
\def\la{\mathrel{\raise.3ex\hbox{$<$\kern-.75em\lower1ex\hbox{$\sim$}}}}

\def\thesection{\arabic{section}}
\def\theequation{\arabic{equation}}
\def\simgt{\mathrel{\raise.3ex\hbox{$>$\kern-.75em\lower1ex\hbox{$\sim$}}}}
\def\simlt{\mathrel{\raise.3ex\hbox{$<$\kern-.75em\lower1ex\hbox{$\sim$}}}}

\newcommand{\nc}{\newcommand}

\nc{\gone}{\bar g_{\pi NN}^{(1)}}
\nc{\gzero}{\bar g_{\pi NN}^{(0)}}
\nc{\al}{\alpha}
\nc{\ga}{\gamma}
\nc{\de}{\delta}
\nc{\ep}{\epsilon}
\nc{\ze}{\zeta}
\nc{\et}{\eta}
\nc{\ka}{\kappa}
\nc{\rh}{\rho}
\nc{\si}{\sigma}
\nc{\ta}{\tau}
\nc{\up}{\upsilon}
\nc{\ph}{\phi}
\nc{\ch}{\chi}
\nc{\ps}{\psi}
\nc{\om}{\omega}
\nc{\Ga}{\Gamma}
\nc{\De}{\Delta}
\nc{\La}{\Lambda}
\nc{\Si}{\Sigma}
\nc{\Up}{\Upsilon}
\nc{\Ph}{\Phi}
\nc{\Ps}{\Psi}
\nc{\Om}{\Omega}
\nc{\ptl}{\partial}
\nc{\del}{\nabla}
\nc{\ov}{\overline}
\nc{\newcaption}[1]{\centerline{\parbox{15cm}{\caption{#1}}}}
\nc{\us}{U(1)$_S$}

\def\beq{\begin{equation}}
\def\eeq{\end{equation}}
\def\bmat{\begin{displaymath}}
\def\emat{\end{displaymath}}
\def\bear{\begin{eqnarray}}
\def\eear{\end{eqnarray}}
\def\ba{\begin{eqnarray}}
\def\ea{\end{eqnarray}}
\def\bery{\begin{array}}
\def\ery{\end{array}}
\def\bit{\begin{itemize}}
\def\eit{\end{itemize}}
\def\ben{\begin{enumerate}}
\def\een{\end{enumerate}}
\def\btab{\begin{tabular}}
\def\etab{\end{tabular}}
\def\btbl{\begin{table}}
\def\etbl{\end{table}}
\def\bfig{\begin{figure}[htb]}
\def\efig{\end{figure}}
\def\bpic{\begin{picture}}
\def\epic{\end{picture}}

\def\nnl{\nonumber \\}

\def\ga{\mathrel{\raise.3ex\hbox{$>$\kern-.75em\lower1ex\hbox{$\sim$}}}}
\def\la{\mathrel{\raise.3ex\hbox{$<$\kern-.75em\lower1ex\hbox{$\sim$}}}}
\def\gappeq{\mathrel{\rlap {\raise.5ex\hbox{$>$}}
{\lower.5ex\hbox{$\sim$}}}}
\def\lappeq{\mathrel{\rlap{\raise.5ex\hbox{$<$}}
{\lower.5ex\hbox{$\sim$}}}}

\def\gyr{{\rm \, G\kern-0.125em yr}}
\def\mev{{\rm \, Me\kern-0.125em V}}
\def\gev{{\rm \, Ge\kern-0.125em V}}
\def\tev{{\rm \, Te\kern-0.125em V}}

\begin{document}

\title{A Closer Look at Dark Vector Splitting Functions in Proton Bremsstrahlung}

\author{S. Foroughi-Abari}
 \email{saeidf@physics.carleton.ca}
\affiliation{Department of Physics, Carleton University,
Ottawa ON K1S 5B6, Canada}
\affiliation{Department of Physics and Astronomy, University of Victoria, Victoria BC V8P 5C2, Canada}
\author{P. Reimitz}
 \email{peter@if.usp.br}
\affiliation{Instituto de F\'{i}sica,
Universidade de Sāo Paulo, 05508-090 Sāo Paulo, SP, Brasil}
\author{A. Ritz}
 \email{aritz@uvic.ca}
\affiliation{Department of Physics and Astronomy, University of Victoria, Victoria BC V8P 5C2, Canada}

\date{September 2024}

\begin{abstract}
\noindent 

High luminosity colliders and fixed target facilities using proton beams are sensitive to new weakly coupled degrees of freedom across a broad mass range. Among the various production modes, bremsstrahlung is particularly important for dark sector degrees of freedom with masses between 0.5 and 2.0 GeV, due to mixing with hadronic resonances. In this paper, we revisit the calculation of dark vector production via initial state radiation in non-single diffractive scattering, using an improved treatment of the splitting functions and timelike electromagnetic form-factors. The approach is benchmarked by applying an analogous calculation to model inclusive $\rho$-meson production, indicating consistency with data from NA27 in the relevant kinematic range.

\end{abstract}
\maketitle

\section{Introduction}

The primary empirical motivations for new physics, in particular the evidence for dark matter and neutrino mass, are relatively agnostic about the mass scale and so have motivated extensive efforts to explore scenarios that are light relative to the weak scale, but necessarily weakly coupled \cite{pospelov2008,Batell:2009yf,Essig:2009nc,Reece:2009un,Freytsis:2009bh,Batell:2009jf,Freytsis:2009ct,Essig:2010xa,Essig:2010gu,McDonald:2010fe,Williams:2011qb,Abrahamyan:2011gv,Archilli:2011zc,Lees:2012ra,Davoudiasl:2012ag,Kahn:2012br,Andreas:2012mt,DS16,CV17,PBC}. This dark sector framework focuses attention on the small number of relevant or marginal interactions, i.e. the scalar, vector and neutrino portals, that could connect an entirely neutral new physics sector to the Standard Model. Notably, the most interesting parameter range in these models is accessible to current and next generation high luminosity collider and fixed target facilities \cite{Batell:2009di,deNiverville:2011it,deNiverville:2012ij,Kahn:2014sra,Adams:2013qkq,Soper:2014ska,Dobrescu:2014ita,Coloma:2015pih,dNCPR,MB1,MB2,Blumlein:2013cua,deNiverville:2016rqh,Feng:2017uoz,Alpigiani:2018fgd,Ariga:2018pin,Dutta:2020vop,Batell:2021blf,Batell:2021aja,Bjorken:2009mm,Izaguirre:2013uxa,Diamond:2013oda,Izaguirre:2014dua,Batell:2014mga,Lees:2017lec,Berlin:2018bsc,NA64:2019imj,Berlin:2020uwy,Krnjaic:2022ozp,Berlin:2018jbm,Bauer:2018onh,Berlin:2023qco,Lu:2023cet,Mongillo:2023hbs,Filimonova:2022pkj,CarrilloGonzalez:2021lxm,Foguel:2022ppx,Batell:2021snh,Garcia:2024uwf}, for example the mediator mass and interaction range required to explain models of thermal relic sub-GeV dark matter. 

The evolving maturity of accelerator-based probes of dark sectors has highlighted the relative advantages of proton and electron beam facilities, with future development in the former case bolstered for example by the developing long and short baseline neutrino physics program at Fermilab~\cite{Machado:2019oxb,DUNE:2020fgq}, and the near-term opportunities for fixed target experiments such as SHiP in the CERN North Area~\cite{SHiP:2015vad}, and forward physics detectors at the HL-LHC~\cite{Feng:2022inv}. This motivates careful analysis of all the relevant hadronic production modes and detection strategies~\cite{Celentano:2020vtu,Capozzi:2021nmp,Blinov:2024pza,LoChiatto:2024guj,Altmannshofer:2022ckw,Curtin:2023bcf,KO24a}. One of the most complex regimes involves the production of dark sector states of 0.5 - 2.0 GeV mass, where enhancement via resonant hadronic mixing is important. This mechanism is particularly advantageous for proton beam facilities where, in the forward region, and given the relatively low momentum transfer, it can be understood as proton bremsstrahlung. Conventional parton-level calculational approaches, e.g. for Drell-Yan, are not currently well-suited for the production of sub-GeV mass states in the far forward region (very low Bjorken $x$). While radiative decays of final state hadrons provide the dominant production mode for dark sector masses below 0.5 GeV, approaches to the 0.5 to 2.0 GeV range have necessarily focussed on coherent radiation from beam protons.

The underlying process of interest involves inclusive production of a dark state via proton beam collisions $p+p/n \rightarrow D+\cdots$ where D is a dark sector state. Such a bremsstrahlung-like process can be characterized via three sub-processes, 
${\rm Brem} \sim {\rm ISR} + {\rm FSR} + {\rm Collective}$, comprising initial and final state radiation and collective effects in the underlying hadronic collision. Our focus here is on initial state radiation (ISR) in proton-proton scattering, as it is well-defined given specified initial states. We revisit data-driven approaches to the calculation of this rate for dark vectors, and specifically kinetically-mixed dark photons, and explore and benchmark these contributions by comparing to data on inclusive rho production. Initial approaches to proton bremsstrahlung generalized the successful Fermi-Weizsacker-Williams (WW) approximation for electron bremsstrahlung \cite{Blumlein:2013cua,deNiverville:2016rqh,Feng:2017uoz}. While this can be straightforwardly applied for quasi-elastic $pp\rightarrow ppV$ radiation, additional assumptions are required to extend this to the most relevant regime of inelastic scattering with a complex final state. In~\cite{Foroughi-Abari:2021zbm}, several approaches to proton bremsstrahlung were studied, including pomeron-exchange models of ISR + FSR for quasi-elastic radiation, the hadronic WW appoximation to quasi-elastic scattering, and an approach following Altarelli-Parisi to ISR in inelastic scattering referred to as the quasi-real approximation (QRA). The latter approach provided a dominant contribution to the total rate, but the approximation leads to some unphysical features in the kinematic distributions, particularly for small vector mass. In this paper, we will address these issues and also provide a more comprehensive analysis of the form-factors at the ISR proton vertex. Our final results for the ISR production rate for dark vectors at colliders with sample beam energies of 400 GeV and 14 TeV are shown in Fig.~\ref{fig:Production}. 

The remaining sections of this paper are organized as follows. Section~\ref{sec:OnShell} discusses the ISR approximation, and the approach to determine a consistent splitting function for the production of a dark vector in proton-proton collisions. We improve on prior work by adopting the Dawson correction to maintain gauge invariance in the massless vector limit. Section~\ref{sec:FormFactors} presents a more precise $ppV$ form-factor in the timelike region, based on data for the Pauli and Dirac proton electromagnetic form factors. The ensuing production rates and kinematic distributions are presented in Section~\ref{sec:production}, along with an application of the QRA approach to inclusive rho production, allowing for a comparison to data as a benchmark. Further technical details on the form-factor parametrizations are covered in an Appendix, with fits provided in an accompanying file for ease of implementation. Section~\ref{sec:Discussion} contains a brief discussion of the results and prospects for further improvements in precision.

%%%%%%%%%%
%%%%%%%%%%
\section{ISR splitting functions}
\label{sec:OnShell}

This analysis will focus on new dark sector degrees of freedom coupled to the Standard Model via the vector portal, and thus the initial state radiation of dark photons $A'$, defined via kinetic mixing with photons, or equivalently their coupling to the electromagnetic current $J_{\rm EM}^\mu$,
\be
 {\cal L} = -\frac{1}{2} \ep F^{\mu\nu} F'_{\mu\nu} = - \ep e J_{\rm EM}^\mu A'_\mu.
\ee
This coupling allows us to infer the coupling of $A'$ to nucleons (and we focus here on protons), which we later parametrize by taking into account both monopole and dipole form-factors, but in this section we retain just the constant charge coupling for simplicity.

Following \cite{Foroughi-Abari:2021zbm}, we use an on-shell approach \cite{Kessler1960SurUM,Baier:1973ms,Baier:1980kx,Nicrosini:1988hw,Altarelli:1977zs}, the quasi-real approximation, to factorize the differential cross section for initial state radiation into a calculable splitting probability and the underlying non-single diffractive proton-proton interaction cross section for which a fit to data is readily available.  

\begin{figure}[t]
    \centering
    \includegraphics[width=0.3\textwidth]{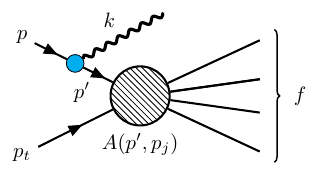}\caption{Dark sector initial state radiation in a generic non diffractive scattering event.}
    \label{fig:ISR}
\end{figure}

The amplitude is represented schematically in Fig.~\ref{fig:ISR}, with a vector radiated from the incoming `beam' proton $p$ with momentum $p_p$, which then undergoes inelastic scattering with the `target' proton $p_t$ with amplitude $A(p'=p-k,p_j)$. Here $k$ is the momentum of the radiated dark vector, and $p_j$ denotes the momenta of the other particles involved in the inelastic process. In the high-energy limit, the quasi-real approximation represents the intermediate proton propagator using an on-shell polarization sum, so that the matrix element takes the form \cite{Foroughi-Abari:2021zbm},
\begin{align}\label{}
&\mathcal{M}_r^{pp_t\rightarrow Vf}(p,k,p_j) \nonumber\\ &\qquad\qquad\approx\sum_{r^{\prime}}\mathcal{M}_{r^{\prime}}^{pp_t\rightarrow f}(p-k,p_j) \,\Big(\frac{V_{r^{\prime}r}}{2k\cdot p-m_V^2}\Big),
\end{align}
with the vertex function  $V_{r^{\prime}r,\lambda} =\ep e \bar{u}^{r^{\prime}}(p^{{\prime}})\cancel{\epsilon}^{\star}_{\lambda}(k)u^r(p)$. 

For concreteness, we now specify the momenta in the infinite momentum frame,
\begin{align}\label{eq:inf-momentum-frame}
    p^\mu&=(p_p+\frac{m_p^2}{2p_p},\mathbf{0},p_p),\\
    k^\mu&=(zp_p+\frac{p_T^2+m_V^2}{2zp_p},\mathbf{p}_T,zp_p)\\
    p'^{\mu}&=((1-z)p_p+\frac{p_T^2+m_p^2}{2p_p(1-z)},-\mathbf{p}_T,(1-z)p_p),
\end{align}
where $z$ is the fraction of the longitudinal momentum carried by the dark photon with transverse momentum $\mathbf{p}_T$.
Integrating over the phase space of the remaining particles in the final state $f$, the ISR cross-section can then be factorized as follows,
\begin{align}\label{diffSplit}
    d\sigma^{pp_t\rightarrow Vf}(s) \approx d\mathcal{P}_{p \rightarrow p^{\prime}V} \times\sigma_{pp}^{\rm NSD}(s^{\prime}).
\end{align}
Here we have introduced the differential splitting probability $d\mathcal{P}_{p\rightarrow p^\prime V}$, and retained only the non-single diffractive (NSD) cross section $\sigma_{pp}^{\rm NSD}(s^{\prime})$, parametrized following Ref.~\cite{Likhoded:2010pc} as
\begin{equation}
\sigma_{\rm NSD}(s)  = 1.76+19.8 \bigg(\frac{s}{\gev^2} \bigg)^{0.057} \quad {\rm mb},
\end{equation}
since radiation in single diffractive processes is suppressed by ISR and FSR interference \cite{Foroughi-Abari:2021zbm}. After accounting for the momentum of the emitted dark vector, the underlying hadronic cross section is a function of $s^{\prime}\simeq s (1-z)$, where this approximation  is valid up to very large $z$ or large angles, beyond which it must be replaced with a complete $p_T$-dependent expression.

The differential splitting probability can be represented in the form,
\begin{align}
d\mathcal{P}_{p \rightarrow p^{\prime}V }
    &
\equiv w(z,p_T^2)dzdp_T^2  
    \nonumber \\
    &
= \bigg(\frac{1}{16\pi^2 z}\frac{|\overline{\mathcal{M}_{p\rightarrow p^{\prime}V}}|^{2}}{\big((p{-}k)^2{-}m_p^2\big)^2}\frac{E_{p^\prime}}{E_p}\bigg)dzdp_T^2,
\label{eq:splitprob}
\end{align}
where $|\overline{\mathcal{M}_{p\rightarrow p^{\prime}V}}|^2 
\equiv \frac{1}{2}\sum_{\rm spin, pol} V_{r^{\prime},r,\lambda} V_{r^{\prime\prime},r,\lambda}^\star$, and we have introduced the splitting function $w(z,p_T^2)$, which will be the primary quantity for the following discussion. 

The validity of this factorized approximation to the ISR process relies on kinematic conditions, including that the off-shell momentum of the intermediate proton should be small relative to scales in the hard scattering, and that the beam energy be the dominant kinematic quantity. As discussed in \cite{Foroughi-Abari:2021zbm}, we require that (i) $(p'^2 - m_p^2) \ll 4(1-z)^2 p^2$, (ii) $p_T,\, m_p \, (m_V) \ll E_p \, (E_k)$, and (iii) introduce an off-shell form factor to control the $pp'V$ vertex as detailed below.

\subsection{Effective QRA splitting function}

Direct calculation, summing over all helicities using the vertex functions above and on-shell polarization vectors, leads to the splitting function \cite{Foroughi-Abari:2021zbm},
\begin{align}\label{SplitVectorMain}
w_{1}(z,p_T^2) &= \frac{\alpha \ep^2}{2\pi H}
\bigg[z-z(1{-}z) \Big(\frac{2m_p^2{+}m_{V}^2}{H}\Big) + \frac{H}{2z m_{V}^2} \bigg] , 
\end{align}
where we have set form factors to unity for the moment, and introduced the kinematic structure function $H(z,p_T^2)\equiv p_T^2+z^2m_p^2+(1-z)m_{V}^2$.

Although this result has a number of the anticipated scaling relations in the relevant kinematic limits, it was noted in \cite{Foroughi-Abari:2021zbm} that it exhibits an unphysical $1/m_V^2$ singularity as $m_V \rightarrow 0$, due to the longitudinal polarization of the vector. This unphysical mass dependence in the splitting function is also observed at the parton level, e.g. within the electroweak $q\rightarrow q'V$ splitting functions used in HERWIG. The $m_V\rightarrow 0$ divergence caused by the longitudinal vector mode was highlighted in \cite{Masouminia:2021kne}, which utilized the effective vector approach of Dawson to remove the part of the longitudinal polarization that is proportional to the four-momentum~\cite{Dawson:1984gx}. We now explore the use of this procedure for the QRA splitting function.

 The contribution related to the longitudinal polarization $\varepsilon_L(k)$ proportional to the four-momenta $k^\mu$ of the massive spin-1 vector gives zero when sandwiched between fermion spinors $\bar{u}^r (p^\prime) \cancel{\varepsilon} u^s(p) \rightarrow \bar{u}^r(p^\prime) \cancel{k} u^s(p) = 0$, where $p$ and $p^\prime = p-k$ denote the four-momenta of the incoming and outgoing fermion, respectively and we used the equation of motion. Thus one can neglect terms proportional to $k$ and it is sufficient to use 
\begin{equation}
    \varepsilon_0^\mu (k) \equiv  \varepsilon^\mu_L(k) - \frac{k^\mu}{m_V} =\frac{m_V}{(k_0+|\vec{k}|)}(-1,\vec{k}/|\vec{k}|).
    \label{eq:eps0}
\end{equation}
The polarization sum, including the transverse contribution $\sum_{i=1,2,L} \varepsilon^{\mu}_i(k) \varepsilon_i^{\nu \star}(k)=-g^{\mu\nu}+\frac{k^{\mu}k^{\nu}}{m_V^2}$,
can then be written using Eq.~(\ref{eq:eps0}) as 
\begin{equation}
    \sum_{i=0,1,2}\varepsilon_i^{\mu}(k) \varepsilon_i^{\nu \star}(k)=-g^{\mu\nu}-\frac{k^\mu\varepsilon^\nu_0 + \varepsilon^\mu_0 k^\nu}{m_V}.
    \label{eq:polsum0}
\end{equation}

In our treatment of the splitting vertex, we consider the outgoing intermediate proton to be on-shell. This necessitates forcing the condition $p^\prime$ to be on-shell in the spinor sandwich such that the intermediate proton's 3-momentum is fixed $ \vec{p^{\prime}}=\vec{p_{p}}- \vec{k} $ while the energy is not automatically conserved at the $ pp^{\prime}V $ vertex, \textit{i.e.} $ E_{p^{\prime}} \neq E_{p}-E_{k}$. This in turn gives $p^\prime = p-k +\delta$ with $\delta^\mu=(H/2z(1{-}z)p_p,\vec{0})$ where we used the momenta defined in the infinite momentum frame Eq.~(\ref{eq:inf-momentum-frame}).

Using the modified polarization sum in Eq.~(\ref{eq:polsum0}), the effective splitting function takes the form
\begin{align}
w^{\rm eff}_{1}(z,p_T^2)&=\frac{\alpha \ep^2}{2\pi H}\bigg[ \frac{1+(1-z)^2}{z} \nonumber \\
 & \qquad -z(1-z)\bigg(\frac{2m_p^2+m_V^2}{H}\bigg)
\bigg] ,
\label{eq:splitfunc2}
\end{align}
which exhibits smooth behaviour in the massless limit and resembles the Altarelli-Parisi splitting kernel of the SM photon. Indeed, this splitting function ensures that the production rate satisfies the scaling relation $d\sigma/dk \propto 1/k$ in the massless vector limit, consistent with the soft photon theorem~\cite{Low:1958sn,Burnett:1967km,DelDuca:1990gz}. In practice, ISR itself is not expected to provide a good approximation for soft radiation in the massless limit, as the quasi-elastic process requires consistent treatment of FSR. Indeed, we have verified that both the pomeron-exchange approach to quasi-elastic $pp\rightarrow ppV$ radiation, and the hadronic WW approach to this quasi-elastic approach, which were discussed in detail in \cite{Foroughi-Abari:2021zbm}, satisfy the soft theorem with the required coefficient. Nonetheless, the approach above proves practical for analyzing radiation of sufficiently massive vectors in inelastic scattering.

The correction leading to the effective splitting function (\ref{eq:splitfunc2}) can be understood as prioritizing gauge invariance at the ISR vertex, but at a cost of slightly violating energy conservation within the QRA scheme. In particular, the impact of dropping terms proportional to $p-p'$ in $\ep_L^\mu$ is a correction to the splitting function $w_1$ that takes the form  $\left(\frac{H}{2zm_V^2} - \frac{2(1{-}z)}{z}\right)$, and naturally diverges as $m_V\rightarrow 0$. This impacts the on-shell condition for the intermediate proton within QRA, but fortunately, energy conservation is controlled by the form factor $F_{pp'V}(p'^2)$ discussed in the next section.

\subsection{Comparison to WW approaches}

The conventional WW approximation as applied successfully to model bremsstrahlung in electron scattering cannot be applied directly to radiation in inelastic scattering, due to the generic nature of the final state. However, a prescription to apply the splitting function obtained in the WW approximation to quasielastic scattering was proposed initially in \cite{Blumlein:2013cua} and adopted in a number of subsequent phenomenological analyses. This approach follows the parametrization of the differential ISR rate 
given in (\ref{diffSplit}), using the total inelastic $pp$ scattering cross section $\si_{pp}^{\rm tot}$ \cite{Blumlein:2013cua},
\begin{align}
    \left(\frac{d\sigma^{\rm inel}_{pp\rightarrow Vf}}{dzdp_T^2}\right)_{\rm Mod\, WW} \equiv w_1^{\rm WW}(z,p_T^2) \times\sigma_{pp}^{\rm tot}(s^{\prime}).
\end{align}
where $w_1^{\rm WW}$ is an effective splitting function that appears within the WW approach to quasi-elastic scattering, with radiation from initial and final state protons,
\begin{align}
    w^{{\rm WW}}_1(z,p_T^2)
    &=\frac{\alpha \ep^2}{2\pi H}\Bigg[\frac{1+(1-z)^2}{z}\nonumber \\
    &-2z(1-z)\left(\frac{2m_p^2+m_V^2}{H}-z^2\frac{2m_p^4}{H^2}\right)\nonumber \\
    &+2z(1-z)\big(1+(1-z)^2 \big)\frac{m_p^2m_V^2}{H^2}\nonumber \\
    &+2z(1-z)^2\frac{m_V^4}{H^2}\Bigg]. \label{BB}
\end{align}
This function will be discussed further below, and we refer to this prescription \cite{Blumlein:2013cua} as the modified WW approach in providing comparisons below.
It is notable that the first two terms in the quasi-elastic splitting function (\ref{BB}) agree with the effective QRA splitting function in (\ref{eq:splitfunc2}) apart from an additional factor of 2 in front of the second term. This difference and the terms quartic in mass in (\ref{BB}) reflect its origin in the two WW sub-processes for ISR + FSR in quasi-elastic scattering \cite{Blumlein:2013cua}.

It is helpful to compare the above approach with a direct calculation of quasi-elastic radiation in elastic scattering, $p+p \rightarrow p+p + V$. This ISR + FSR combination was computed in \cite{Foroughi-Abari:2021zbm} via both an explicit calculation using pomeron exchange, and a straightforward hadronic generalization of the WW approximation used for electron beams, with both approaches agreeing well (see also \cite{Gorbunov:2023jnx} for a recent analysis). Importantly, it was observed that the ISR +FSR combination showed substantial interference with the resulting rate being substantially lower than either ISR or FSR alone. Due to the more complex final state, this interference is not expected to apply to ISR for non-single diffractive scattering.  As the quasi-elastic rate combines both ISR + FSR, we simply define an effective splitting function as follows,
\be
 w^{\rm elastic}(z, p_T^2,s) \equiv \frac{1}{\si_{pp}^{\rm el}(s)}\frac{d\si^{\rm el}_{pp\rightarrow ppV}}{dz dp_T^2},
 \label{elastic}
\ee 
where $\si_{pp}^{\rm el}$ is the elastic scattering cross section,
and $w^{\rm elastic}$ can be compared numerically to the ISR splitting functions. 

To explore this further, we note that the hadronic WW approach provides a good approximation to (\ref{elastic}) in this regime and can be represented analytically in terms of the splitting function $w_1^{{\rm WW}}$ \cite{Foroughi-Abari:2021zbm},
\begin{align}\label{WW}
   \bigg(\frac{d \sigma^{\rm el}_{pp\rightarrow ppV}}{dz dp_T^2}\bigg)_{\mathrm{WW}} 
    \approx w^{{\rm WW}}_1 (z,p_T^2) \frac{z^2}{H}\bigg(\frac{\chi}{4\pi}\bigg). 
\end{align}
where $\chi$ reflects the factorized dependence on the underlying elastic scattering cross section,
\begin{equation}
    \frac{\chi}{4\pi} = \int_{t_{\mathrm{min}}}^{t_{\mathrm{max}}} dt (t{-}t_{\mathrm{min}}) \frac{d \sigma^{\rm el}_{pp}}{dt},
    \label{chi}
\end{equation}
with $t_{\rm min}{\approx}-H^2/(2z(1{-}z)p_p)^2$ and $t_{\rm max}{\approx}-2(1{-}z)m_p p_p$. The WW splitting function is defined in this case by $w^{{\rm WW}}_1(z,p_T^2){=}(\alpha \ep^2/2\pi H) (1{-}z)/2z \mathcal{A}_{t{=}t_{\text {min }}}^{2 \rightarrow 2}$
and can be re-expressed in the form given in (\ref{BB}), using the expression $\mathcal{A}_{t{=}t_{\text {min }}}^{2 \rightarrow 2} {=}2 (1{+}(1{-}z)^2)/(1{-}z){-}4 z^2p_T^2(2m_p^2{+}m_V^2)/H^2$
for the squared amplitude for the WW 2 to 2 sub-process of 2 to 3 quasi-elastic scattering~\cite{Liu:2017htz}.

It is notable that while both the direct and WW computations of this quasielastic scattering rate in \cite{Foroughi-Abari:2021zbm} utilized the model of pomeron exchange for the underlying process of proton-proton scattering, the splitting probability itself is quite robust to the choice of underlying interaction model. As shown recently in \cite{Gorbunov:2023jnx}, an analogous WW approach, that instead utilizes massless vector propagators and vertices leads to an almost identical result.

We will compare and contrast the differential distributions obtained using the effective QRA and modified WW approaches to ISR along with radiation in quasi-elastic scattering after discussing the important role played by form factors.

%%%%%%%%%%%%%%%
%%%%%%%%%%%%%%%
\section{Form Factors}\label{sec:FormFactors}

Coherent emission of a dark vector with timelike momentum from a composite state such as a proton requires the introduction of form factors that depend on at least two invariants $k^2$ and $p'^2$ due to the intermediate proton nominally being off-shell. In practice, the QRA formalism requires that $p'$ not be too far off-shell for overall energy conservation. We will parametrize the form-factors as follows \cite{Foroughi-Abari:2021zbm},
\be
 F_i(t,p'^2) = {\cal K}_{pp'V}(p'^2) \times F_i(t)
\ee
where $F_i(t=k^2)$, for $i=1,2$, denote the conventional on-shell electromagnetic form factors for the proton, while ${\cal K}_{pp'V}$ controls the off-shell behaviour of the intermediate proton line. This function plays an important role within the QRA approach in filtering out potentially unphysical kinematic contributions due to the lack of precise energy conservation at the dark vector vertex. This implies that ${\cal K}$ has the role of a `kinematic filter' within the QRA formalism, that goes beyond the normal expectation of simply accounting for compositeness in the off-shell proton. For this reason, rather than pursuing a more physical approach, adopted e.g. by Davidson and Workman \cite{Davidson:2001rk,Penner:2002md} where issues of gauge invariance can also be addressed, we model ${\cal K}$ via a simple dipole form \cite{Foroughi-Abari:2021zbm},
\be
{\cal K}_{pp'V} (p'^2) = \frac{1}{1+(p'^2-m_p^2)^2/\Lambda_p^4},
\ee
with a sliding scale $\La_p$ controlling the level of off-shell contributions. The impact of this form-factor is most easily understood by considering the rest frame of the emitting beam proton $p \rightarrow p' + V$. The form factor limits the off-shellness of $p'$, so that all kinematic scales are small ($\sim m_p$) in this frame, while in the lab frame the momentum of $V$ approaches $E_{\rm beam}/2$. Thus, in considering small $V$ momentum, $p'$ is necessarily more off-shell and the form factor suppresses the rate. This effect becomes less significant as the $V$ mass is lowered below the hadronic scale, and indeed the effect of the form-factor is negligible for radiating soft (massless) photons.

Turning now to the electromagnetic form factors $F_i(t)$, we extend earlier work by defining the coupling to nucleons in the standard manner taking into account both electric
(Coulomb) and magnetic (spin-flip) interactions,
\begin{align}
\langle N|J_{\rm em}^\mu|N \rangle 
&{=}\bar{u}(p^\prime) \bigg[ \gamma^\mu F^N_1(t) \\
&\qquad\qquad {+} \frac{i\sigma^{\mu\nu}(p^\prime{-}p)_\nu}{2m_N} F^N_2(t) \bigg] u(p),  \nonumber
\end{align}
where $k=p-p'$ and $t=k^2$ is timelike for the ISR vertex. 
The $ p\rightarrow p^{\prime}V$ vertex function is then generalized accordingly, and evaluating $|\overline{\mathcal{M}_{p\rightarrow p^{\prime}V}}|^2$ with the Dawson-corrected polarization of Eq. (\ref{eq:polsum0}) we obtain

\begin{align}
w_F(z,p_T^2) &= w_1^{\rm eff}(z,p_T^2)\big|F_1(k^2,p'^2)\big|^2 \\
&\;\;\; 
 + w_2^{\rm eff}(z,p_T^2)\big|F_2(k^2,p'^2)\big|^2 \nonumber\\
& \;\;\;
+w^{\rm eff}_{12}(z,p_T^2) \operatorname{Re}\big[F_1(k^2,p'^2)F_2^\star(k^2,p'^2)\big], \nonumber
\label{eq:splitfuncformfactor}
\end{align}
where $w_1^{\rm eff}$ is the splitting function given earlier in (\ref{eq:splitfunc2}), and
\begin{align}
 & w_2^{\rm eff}(z,p_T^2) - w_1^{\rm eff}(z,p_T^2) \nonumber\\
 & \qquad\qquad =  \frac{\alpha\epsilon^2}{4\pi m_p^2}\bigg[ 1 - (1{-}z)\bigg(\frac{4m_p^2{+}m_V^2}{H}\bigg) \bigg] \nonumber\\
 & \qquad\qquad\qquad\qquad \times\bigg[ \frac{1}{z}-\frac{1}{4} z\big(\frac{4m_p^2{-}m_V^2}{H}\big) \bigg],
 \nonumber \\ 
& w^{\rm eff}_{12}(z,p_T^2) = \frac{\alpha\epsilon^2}{2\pi H}\bigg[ 2z - z(1{-}z)\frac{3m_V^2}{H} \bigg].
\end{align}

Information about the Sachs parametrization of the form factors $G_E(t)=F_1(t)+t/(4m^2) F_2(t)$ and $G_M(t)=F_1(t)+F_2(t)$ (and thus the Dirac and Pauli form factors) is primarily obtained through measurements of \(\sigma(e^+e^- \rightarrow N\bar{N})\) in the physical time-like region and from \(e^-p \rightarrow e^-p\) elastic scattering in the space-like region. Subsequently, utilizing the normalization from the electric charge and magnetic moment and large $t$ asymptotics from perturbative QCD (quark counting rules) as guidelines, an extended Vector Meson Dominance (VMD) model is employed to extrapolate further into the so-called unphysical region from $t=0$ to $t=4m_N^2$. Ref.~\cite{Faessler:2009tn} utilized a minimal VMD model for the form factors with only \(\rho\) and \(\omega\) resonances, assuming identical masses for ground states (\(m_\rho = m_\omega = 0.770\) GeV) as well as for the excited states. In this paper, we explore a more elaborate resonance-based model, along with a modern dispersive approach for parametrizing nucleon electromagnetic form factors. The proton and neutron Pauli and Dirac form factors in the time-like region, as predicted by these models, are compared in Appendix~\ref{app:EM_FF}. 

\begin{figure}[t]
\centering
\includegraphics[width=0.48\textwidth]{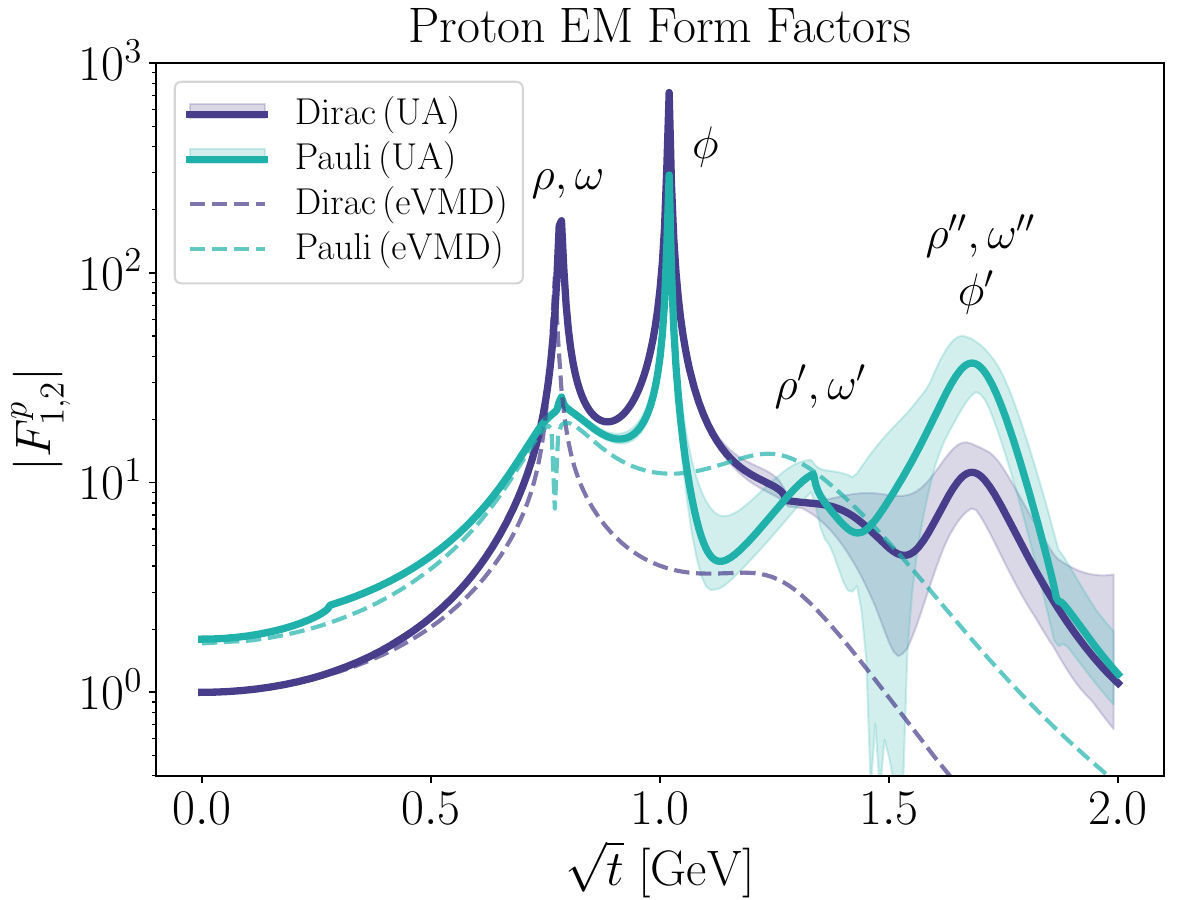}
    \caption{Proton electromagnetic form factors in the time-like region based on the Unitary and Analytic (UA) model~\cite{Adamuscin:2016rer}, with fits provided in an accompanying files \cite{fitfiles}. For comparison, dashed lines represent the simpler extended Vector Meson Dominance (eVMD) model~\cite{Faessler:2009tn,deNiverville:2016rqh}. The meson labels indicate the approximate locations of the resonance structures, with $^\prime$ and $^{\prime\prime}$ denoting the first and second excited resonance states, respectively. See the text for further details.}
    \label{fig:EMFormFactor}
\end{figure}

We observe that the inclusion of a $\phi$ pole and continuum contributions to the isoscalar part of the form factor may have a sizable impact on dark photon production predictions for masses in the GeV range. Although we expect direct $\phi$-nucleon interactions to be highly suppressed according to the OZI rule~\cite{Okubo:1963fa,Zweig:1964jf,Iizuka:1966fk}, higher-order effects generate a non-negligible effective $\phi NN$ interaction and an associated enhancement in the form-factor around $\sqrt{t} \sim m_\phi\sim 1$ GeV. Indeed, as discussed in the dispersive analysis of~\cite{Lin:2021xrc}, the isoscalar region around $\sim 1$~GeV consists primarily of  $\rho\pi$ and $K\bar{K}$ continuum contributions in addition to an intrinsic $\phi-$meson pole that is damped due to cancelations. In contrast, the $\omega$ resonance is well isolated and prominent, making it possible to extract an $\omega NN$ coupling. Alternatively, the Unitary-Analytic (UA) model~\cite{Adamuscin:2016rer} which generalizes the approach of \cite{Faessler:2009tn} contains just a series of neutral vector-meson poles within the framework of vector meson dominance (VMD)~\cite{Sakurai:1960ju,Kroll:1967it,Lee:1967iv}. This approach entails a clear peak structure in the form factors and gives much more significance to the $\phi$ pole as seen in Fig.~\ref{fig:EMFormFactor}. 
However, in this model, the overall height of the form-factor in the relevant timelike region is somewhat lower than the dispersive result, as seen in Fig.~\ref{fig:EM_FF_fit}. Thus, to be conservative we adopt the UA model to determine physical rates. Nonetheless, we note that both modeling approaches lead to a broad enhancement of the form factors above 1~GeV, relative to the simplified VMD model, due to the impact of broad resonances.

There are several possibilities to estimate the form factor uncertainty in the 'unphysical region'. The number of data points outside the unphysical region ensures that the uncertainty in the fit itself is rather small, but this does not account for the intrinsic uncertainty in the model which is more important here as we wish to use the form factors in the unphysical timelike region. Indeed, the fit values are expected to be primarily sensitive to the region where the fit function is actually fitted to data. However, our interest lies in the hadronic resonance region with parameters fixed to their most recent PDG values~\cite{Workman:2022ynf}. Consequently, we focus on the intrinsic uncertainty in choosing the fit function parametrization and vary the resonance masses around their documented experimental uncertainty. We observe that the additional uncertainty in the widths, in particular of heavier resonances, inflates the uncertainty bands significantly further and limits the confidence in the fit in that region (as indicated by dashed lines in our final production rates). Our final results for the form factors, utilizing the UA model are shown in Fig.~\ref{fig:EMFormFactor}. Further details can be found in Appendix~\ref{app:EM_FF}. 

%%%%%%%%%%%
\section{Production rates and benchmarks}\label{sec:production}

\begin{figure*}[t]
\centering
\includegraphics[width=0.31\textwidth]{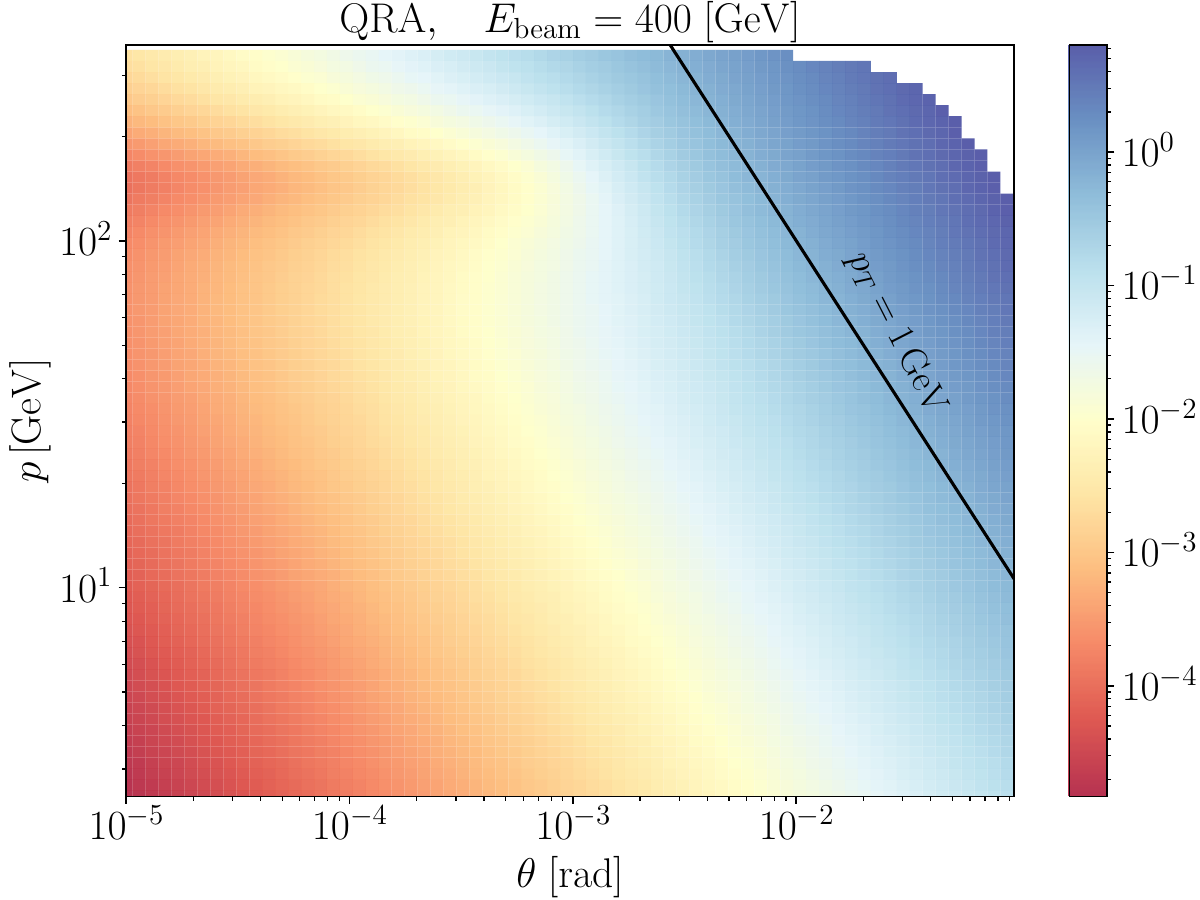}
\includegraphics[width=0.31\textwidth]{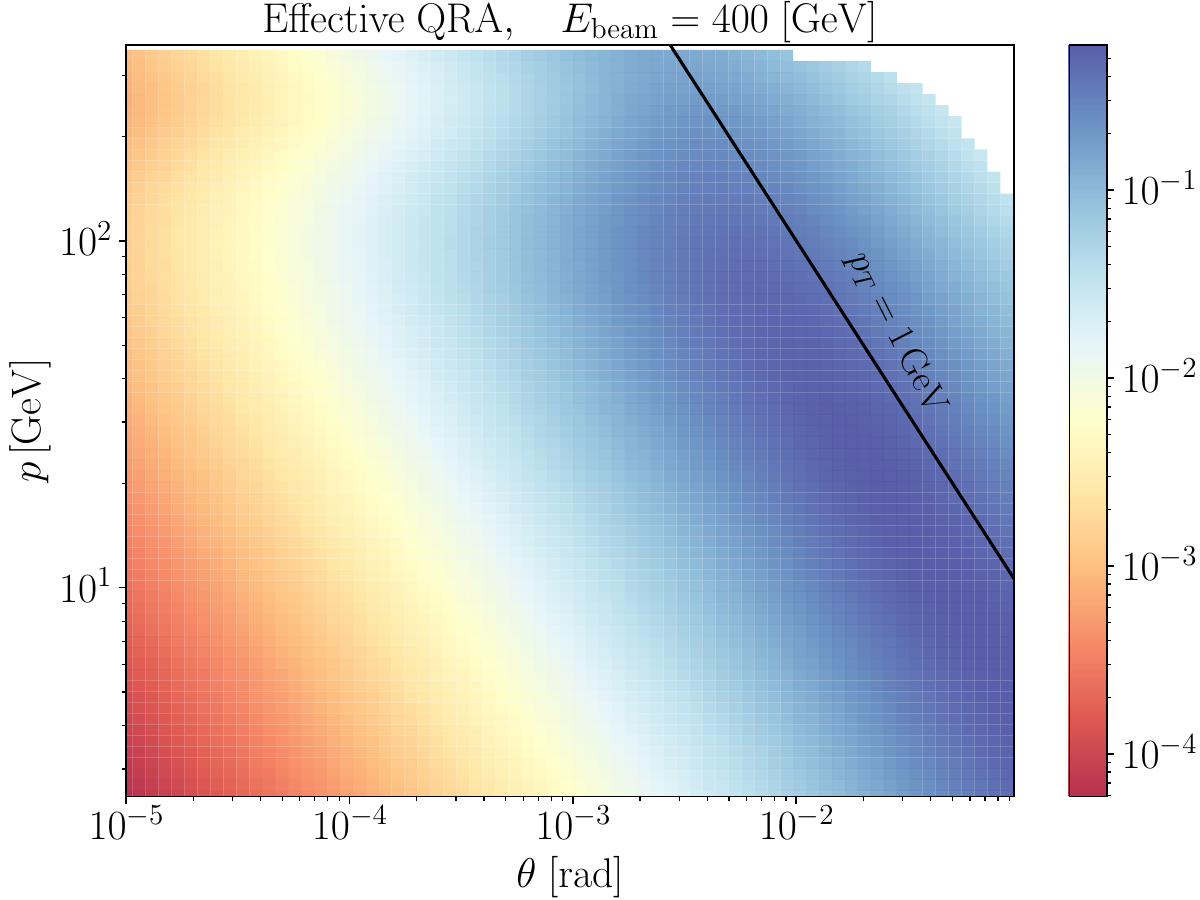}
\includegraphics[width=0.31\textwidth]{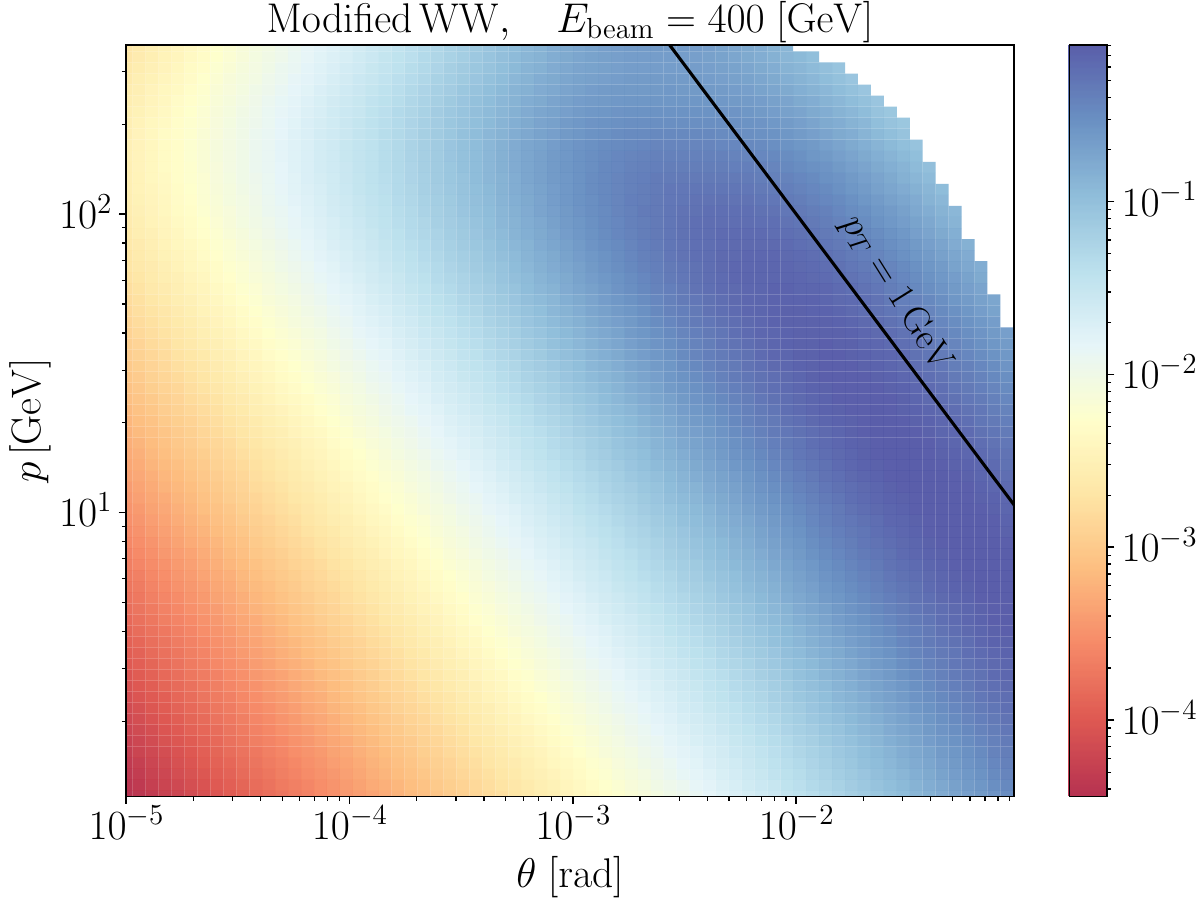}
    \\
    \vspace{0.1cm}
\includegraphics[width=0.31\textwidth]{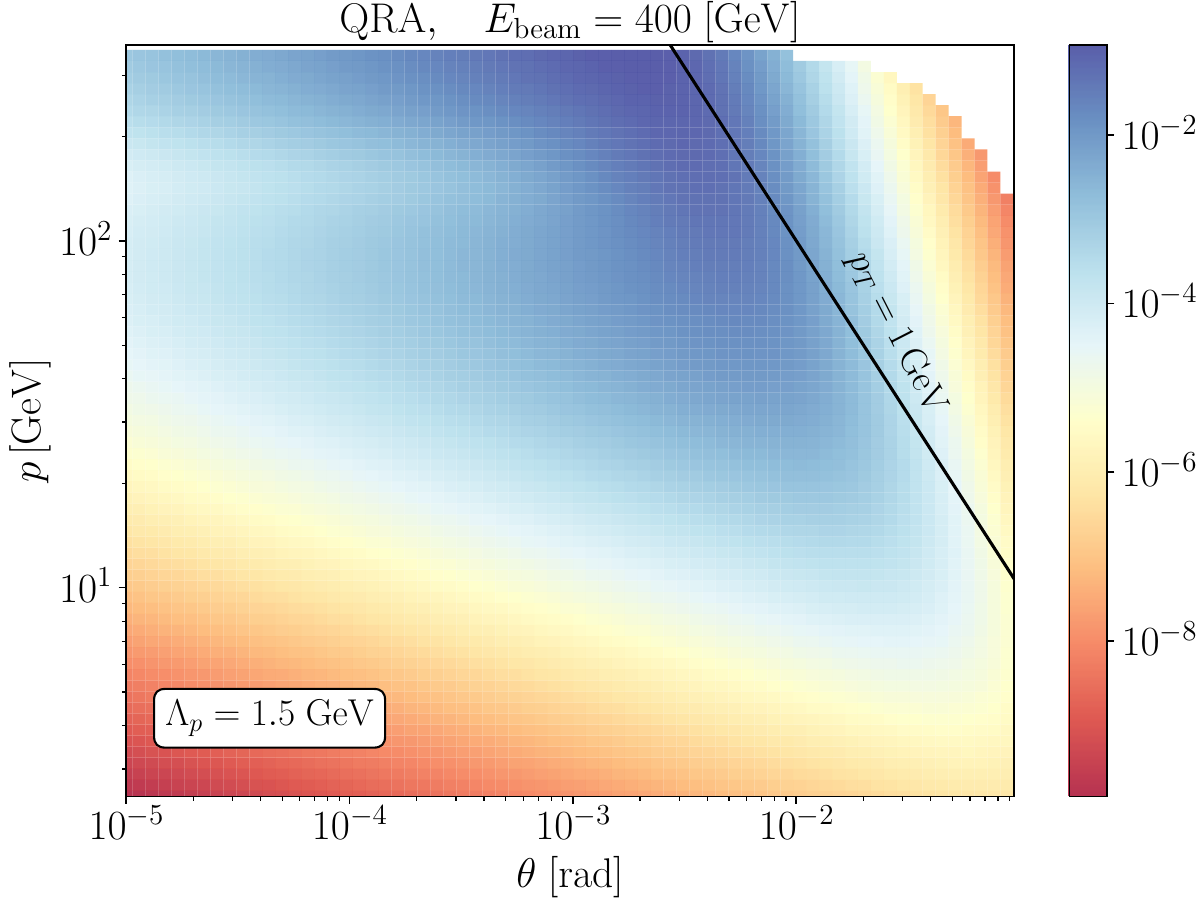}
\includegraphics[width=0.31\textwidth]{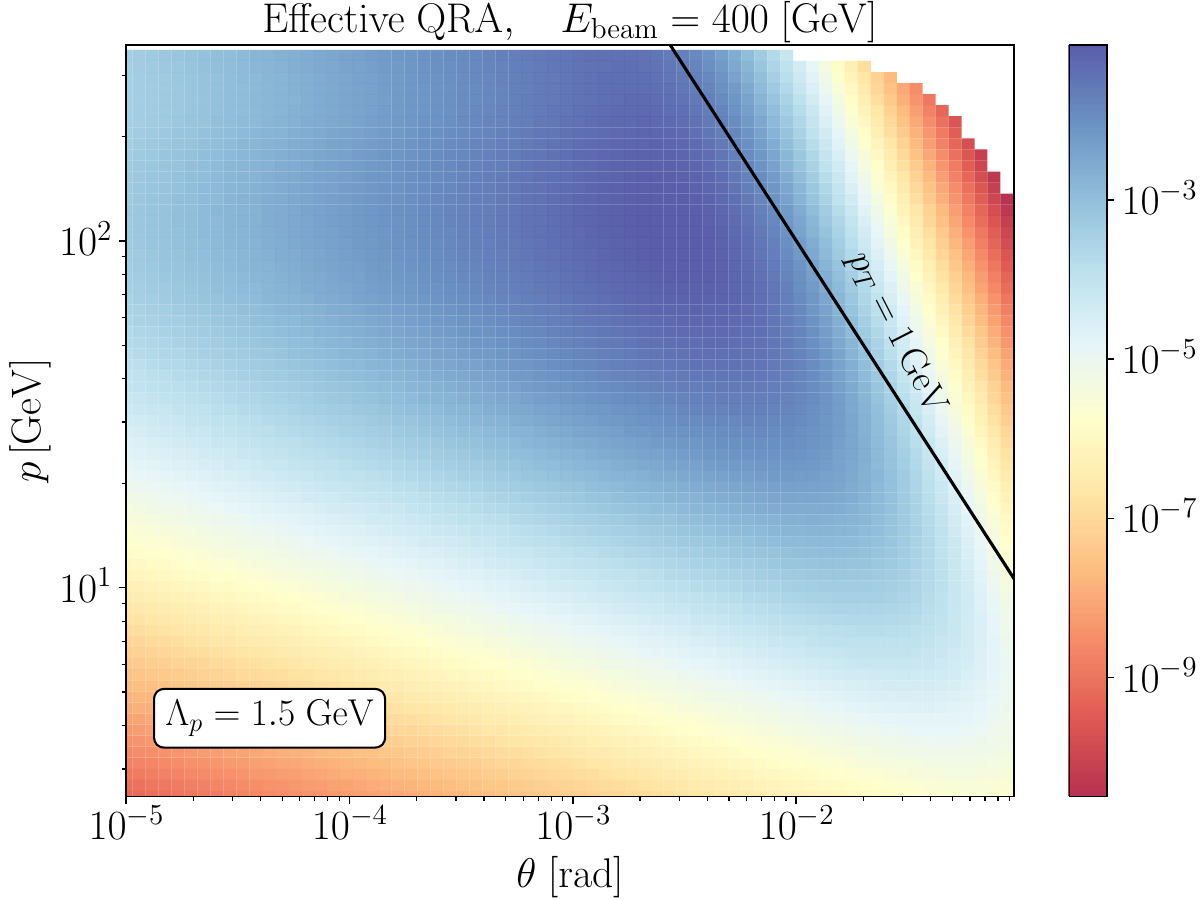}
\includegraphics[width=0.31\textwidth]{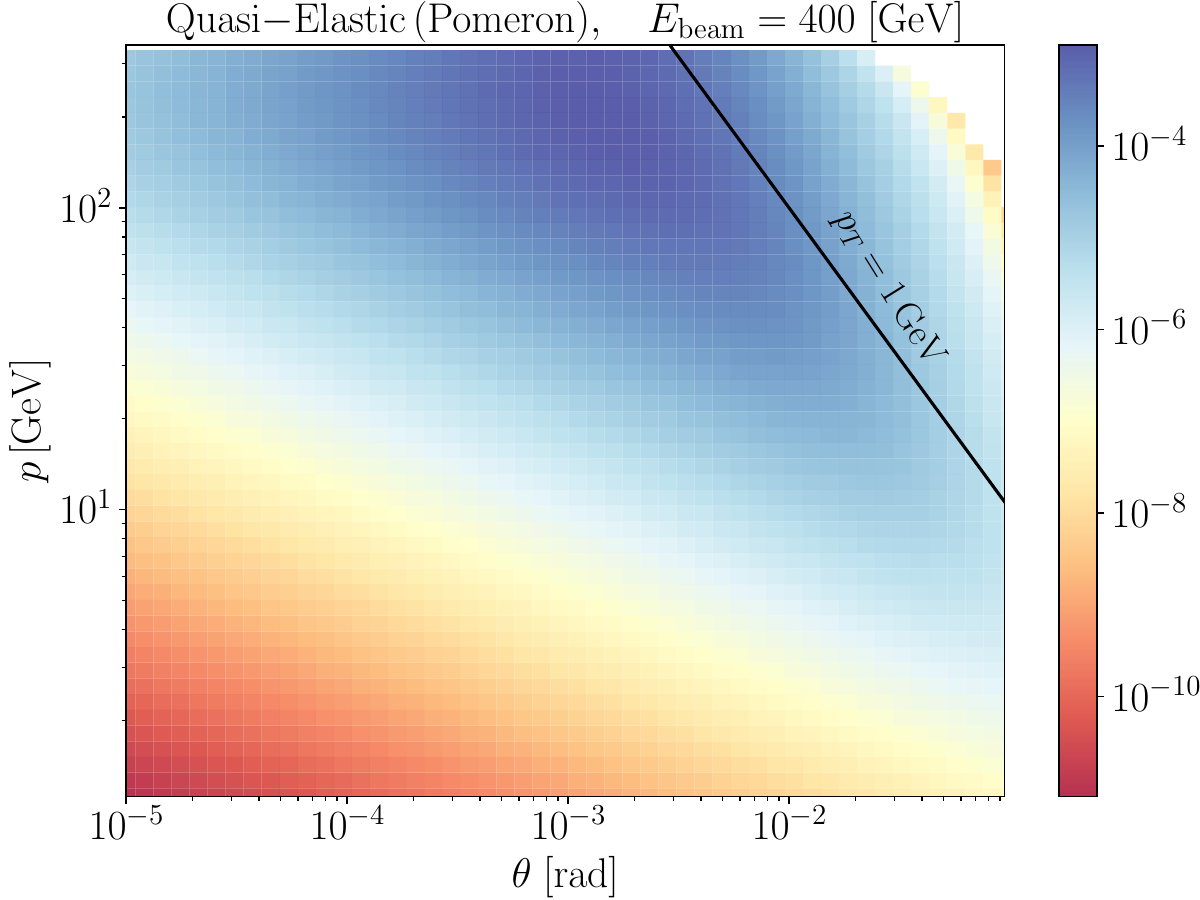}
    \caption{Production distributions $d\mathcal{P}_{p \rightarrow p^{\prime}V }$ for $m_V = 0.5 ~\rm{GeV}$ at a 400 GeV fixed target experiment are shown for several approximations to proton bremsstrahlung, labeled QRA, effective QRA, modified WW, and quasi-elastic, as described in the text. The plots in the first two columns show the impact of the off-shell form-factor ${\cal K}$ within the QRA and effective QRA approaches.}
    \label{fig:QRA}
\end{figure*}

In this section, we present the final production rates and distributions using the effective QRA approximation for ISR, along with comparisons to other approaches. We also test the effectiveness of this approach within the Standard Model by using the corresponding ISR vertex with an off-shell photon to model the production of $\rho$-mesons via proton bremsstrahlung. This allows a direct comparison to inclusive $\rho$-production data from NA27 to benchmark the rate. 

\subsection{Dark vector production rates and distributions}

\begin{figure*}[t]
\centering
\includegraphics[width=0.31\textwidth]{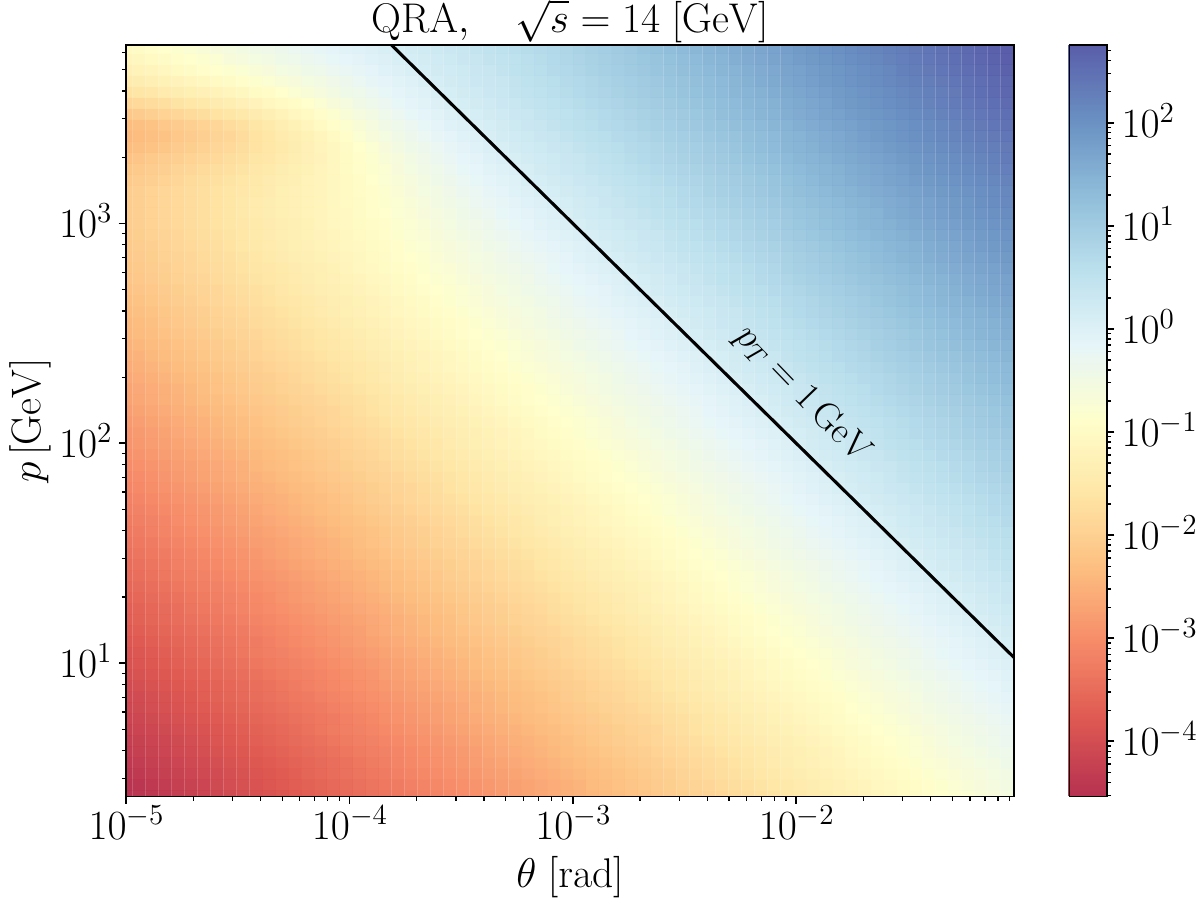}
\includegraphics[width=0.31\textwidth]{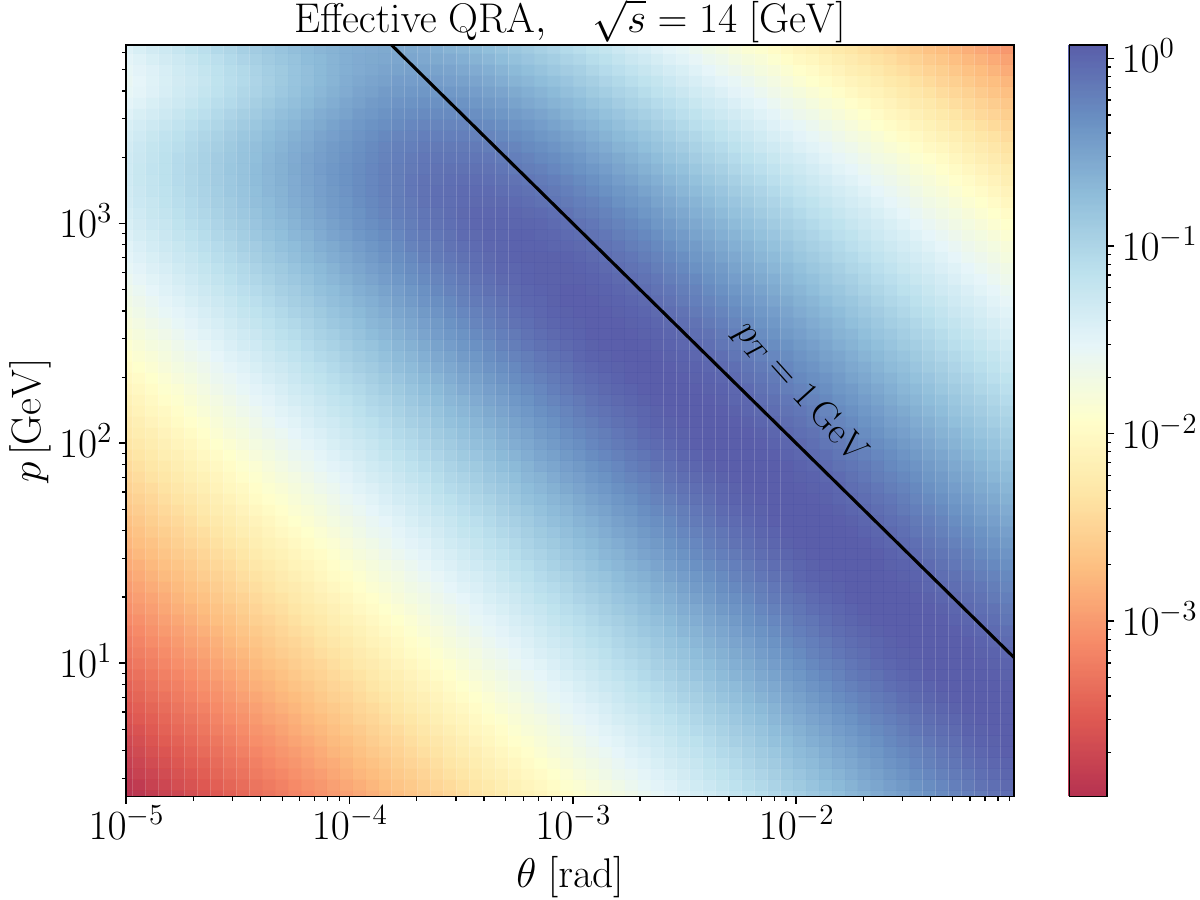}
\includegraphics[width=0.31\textwidth]{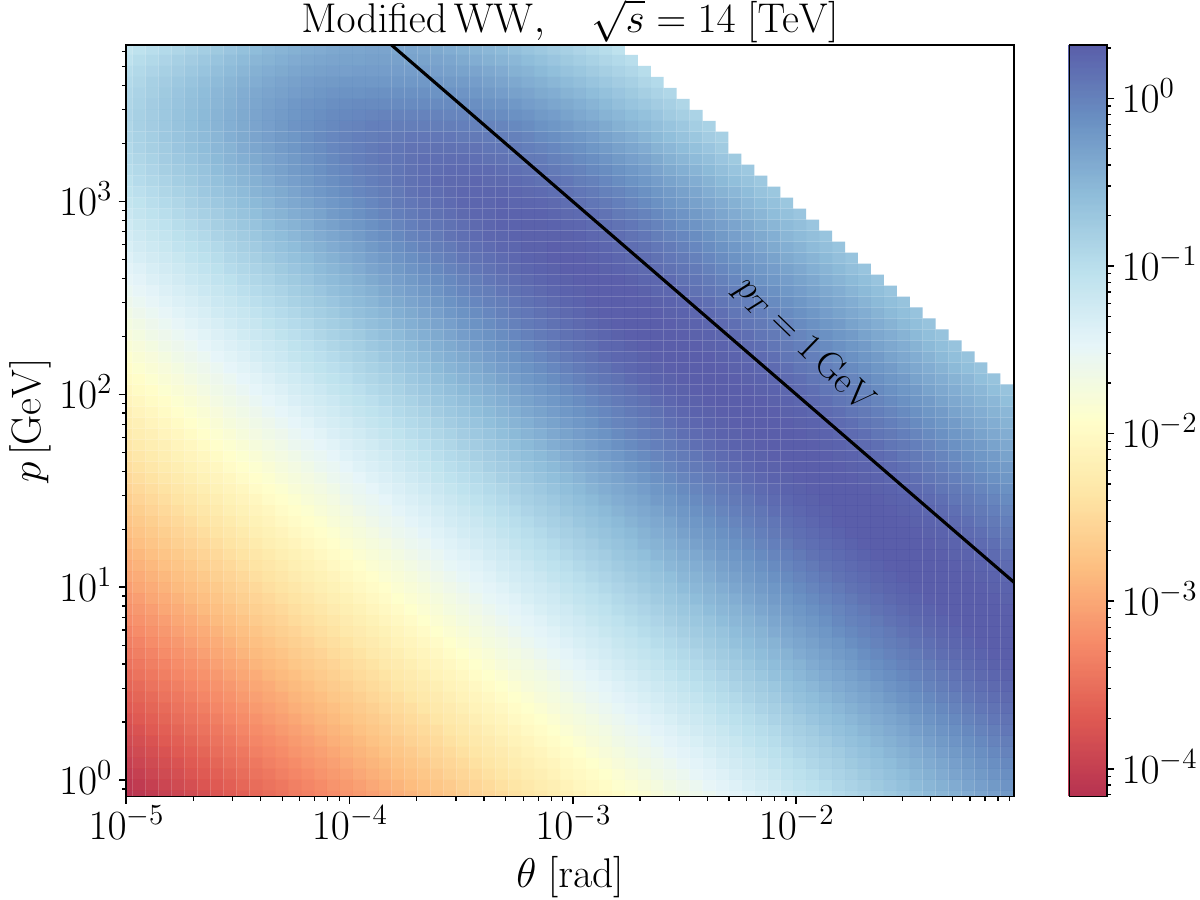}
    \\
    \vspace{0.1cm}
\includegraphics[width=0.31\textwidth]{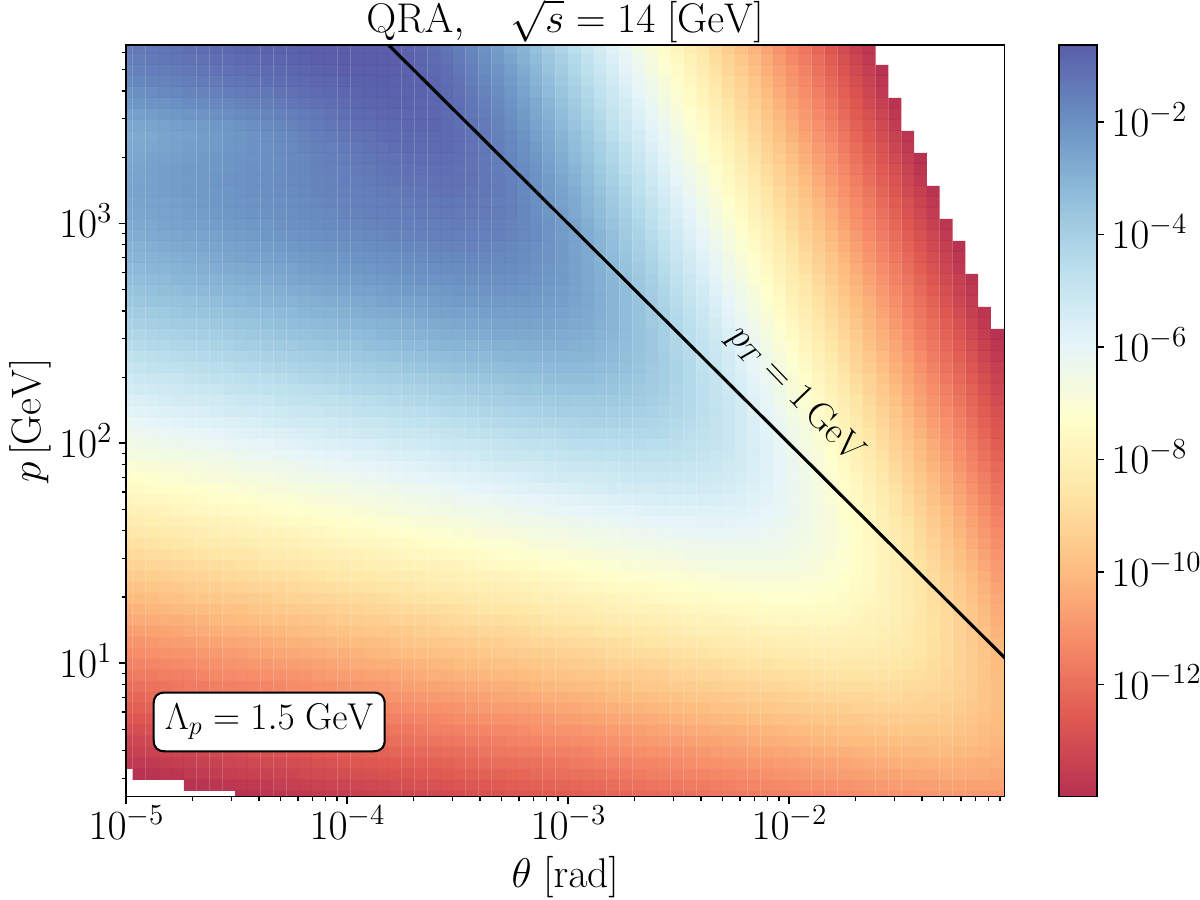}
\includegraphics[width=0.31\textwidth]{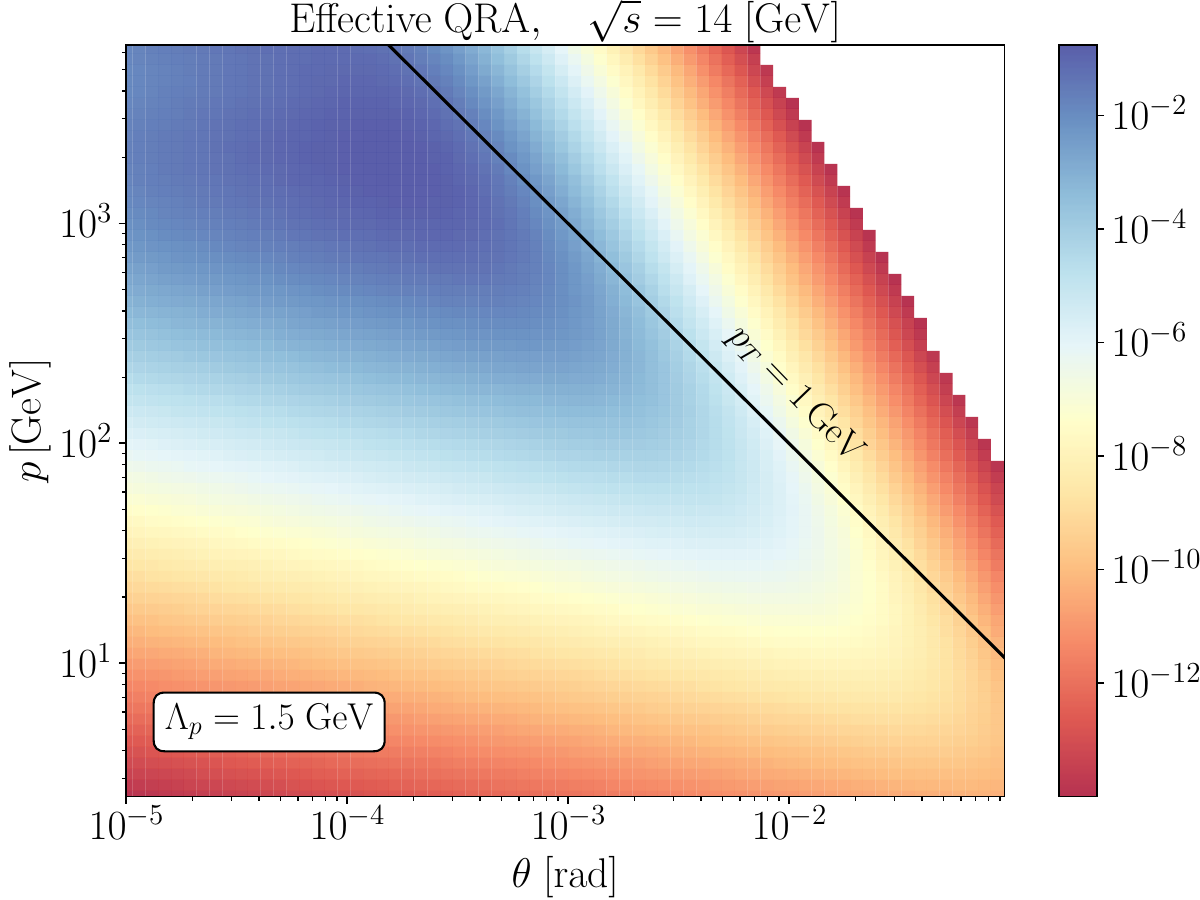}
\includegraphics[width=0.31\textwidth]{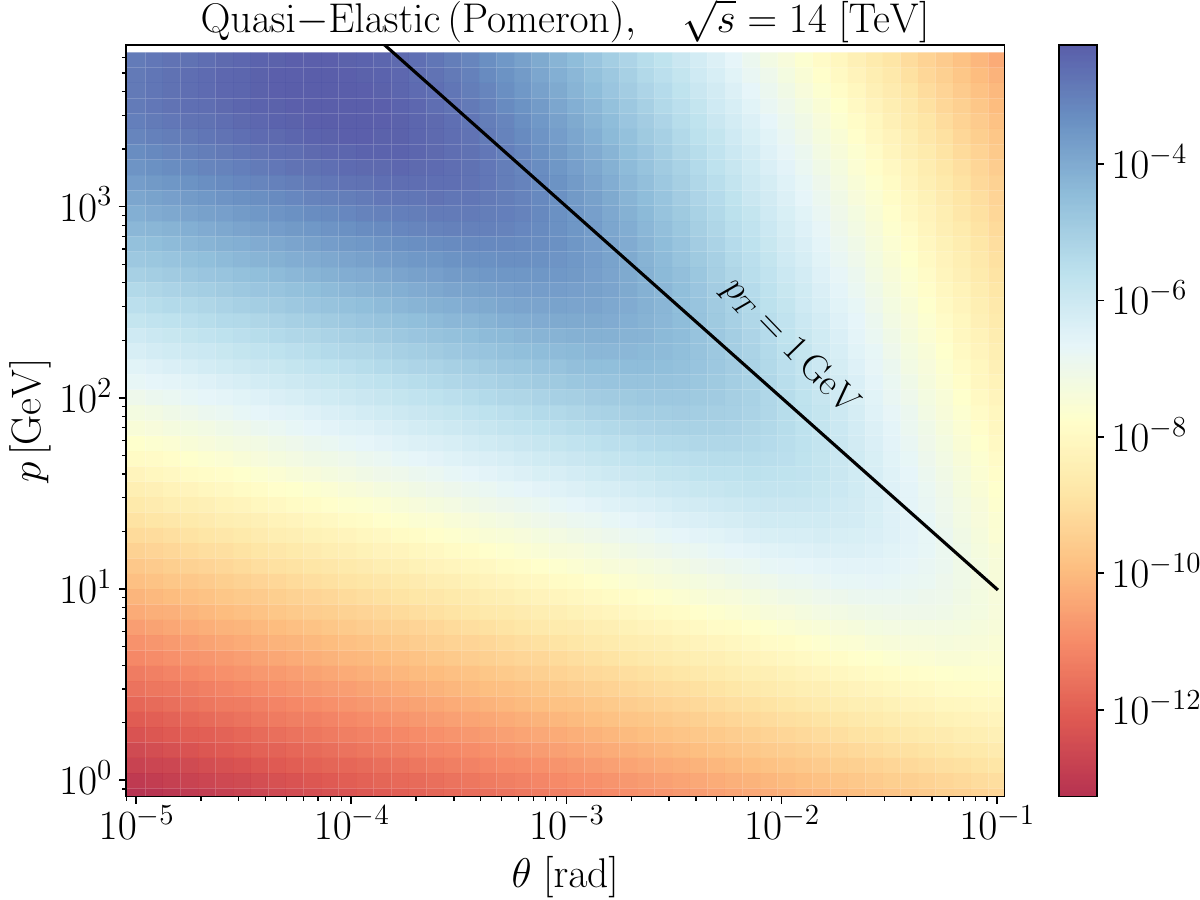}
    \caption{Production distributions for $m_V = 0.5 ~\rm{GeV}$ in the centre-of-mass frame at a forward LHC or HL-LHC experiment are shown for a variety of approximations to proton bremsstrahlung, following the same layout as Fig.~\ref{fig:QRA}.}
    \label{fig:QRA_CM}
\end{figure*}

We first illustrate the various differential distributions $d\mathcal{P}_{p \rightarrow p^{\prime}V }$ for two different beam energies in Figs.~\ref{fig:QRA} and \ref{fig:QRA_CM}. These panels help to clarify the impact of the off-shell form-factor ${\cal K}$, while the EM form-factors are not included to aid these comparisons.

It is instructive to compare the QRA approach to ISR with the well-defined computation for quasi-elastic scattering, which for example satisfies the correct scaling in the soft photon limit. It is apparent that the rate of the quasi-elastic process is suppressed when off-shellness is large ($H/z \gg 1$). We observe that this feature is not apparent in the naive rates obtained using either QRA or the modified WW approximation. However, this feature is restored by including the off-shell form-factor ${\cal K}$ for the intermediate proton line. Indeed, when the energy of the radiated massive state is low ($z \rightarrow 0$, $m_V \sim m_p$), the off-shell form factor applied to ISR scales as $z^2$ providing a corresponding suppression factor. A more precise comparison follows from the hadronic WW approximation to the quasi-elastic rate, as given in (\ref{WW}), $d\si/dz dp_T^2 \sim z^2 \times w_1^{\rm{WW}} \times \chi / (4\pi H)$. This can be approximated further in the regime where $H \sim m_p^2$, and to evaluate the integral expression for $\chi$, we approximate the elastic cross-section by a simple exponential fall-off $d\sigma^{\rm el}_{pp}/dt \propto \exp(-B|t|)$, where the diffractive slope $B(s)\sim \mathcal{O}(10) \, \rm{mb}$~\cite{TOTEM:2017asr}. This gives $ \chi /4\pi \approx \sigma^{\rm el}_{pp}(s)/B(s) \sim \mathcal{O}(1)$ for $B|t_{\rm min}| \ll 1,$ so the scaling of the rate with $z^2$ is apparent. This kinematic feature helps to explain why the ISR distributions, when modulated by the off-shell form factor exhibit similar functional form to that for quasi-elastic scattering.

Another notable feature of the distributions is that the incorporation of the Dawson correction into the effective QRA approach removes an unphysical feature for small angles and large momenta, that is observable in the QRA distributions. Overall, the consistency of the effective QRA distribution with that derived for quasi-elastic scattering adds confidence to the consistency of the approach.

\begin{figure*}[t]
\centering
\includegraphics[width=0.49\textwidth]{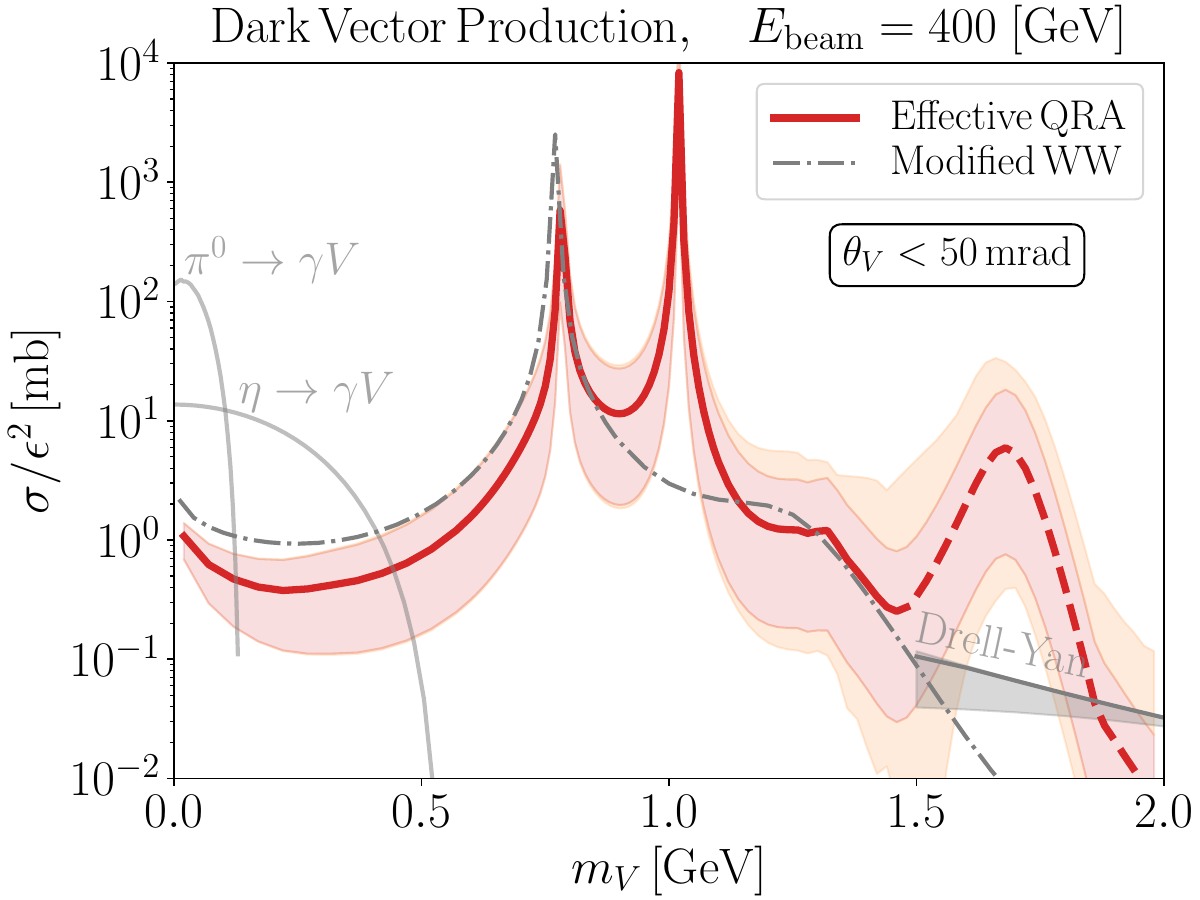}
\includegraphics[width=0.49\textwidth]{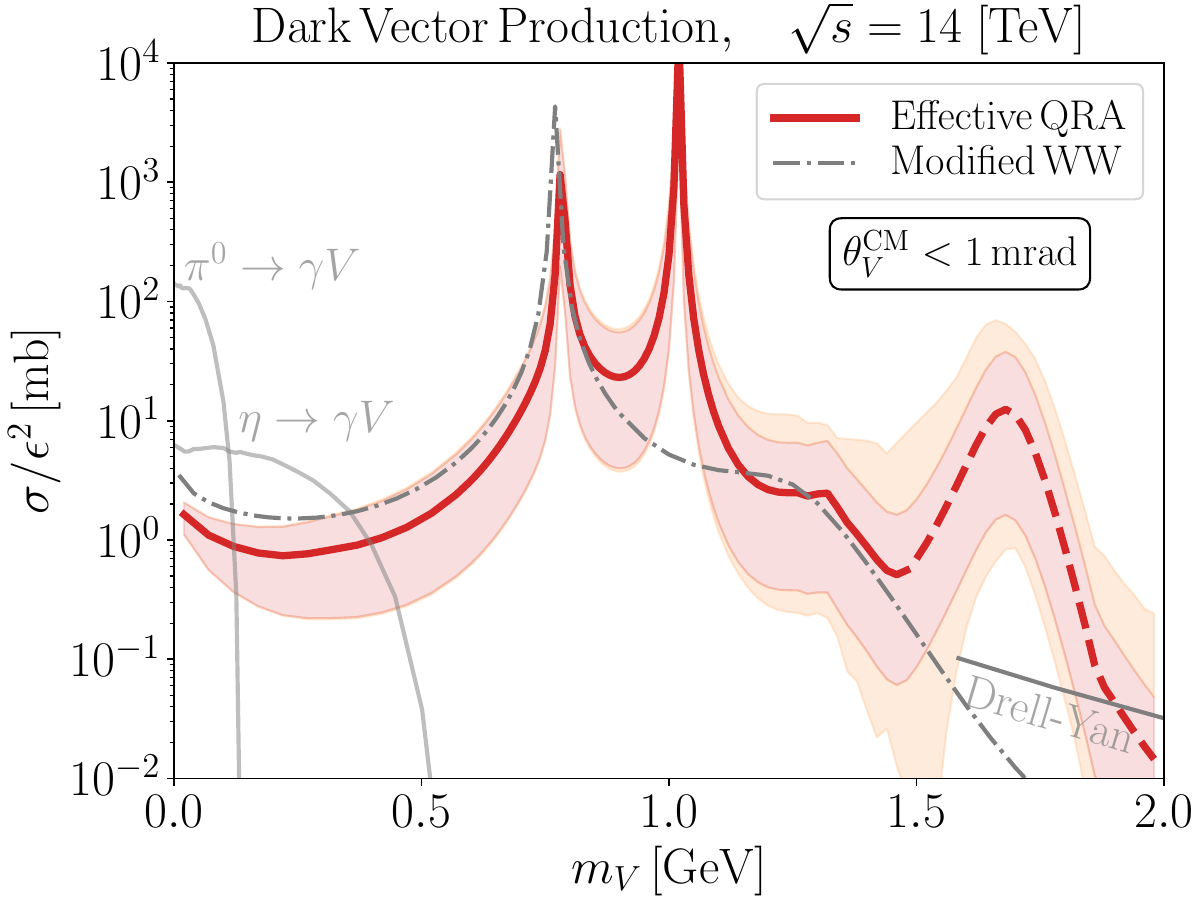}
    \caption{The production cross-section of dark vectors is shown as a function of mass for (\textbf{left}) a 400 GeV fixed target beam, with an angular range of 50 mrad from the beam axis, and (\textbf{right}) at $14 \tev$ (center of mass) energy, within 1 mrad of the beam axis (center of mass frame). The improved quasi-real approximation in non-single diffractive scattering with a cut-off $\Lambda_p=1.5 \gev$ is shown in red. The red uncertainty band is represented by varying $\Lambda_p$ within the interval $[1, 2] \gev$. Stacked on top in orange is an estimate of the uncertainty in the EM form factor parametrization. The dashed grey curve employs the modified WW approach~\cite{Blumlein:2013cua}, utilizing the simpler VMD form factor, with a cut on transverse momentum $p_T < 1 \gev$~\cite{Berlin:2018jbm}. Lighter grey curves depict other production channels, as labeled.}
    \vspace{-0.5cm}
    \label{fig:Production}
\end{figure*}

With the distributions in hand, on incorporating the timelike EM form-factors, we can determine the total production rate for vectors of a given mass, as shown in Fig.~\ref{fig:Production}. This figure indicates the impact of resonant mixing with vector mesons, and the uncertainty bands are determined by varying the filter scale $\La_p$, along with the uncertainty in the EM form factors. We emphasize that there are additional production modes, e.g. final state radiation, that are not yet incorporated into the modeling of dark vector rates.

%%%%%%%%%%%%%%
%%%%%%%%%%%%%%

%%%%%%%%%%%%%%
\subsection{Comparison to inclusive \texorpdfstring{$\rho$}--meson production}

We now consider benchmarking the ISR differential rate in relation to data for inclusive light vector meson production, focussing on $\rho (770)$.

Fits to the kinematic distributions observed in inclusive meson production, including the vector mesons, $\rho,\omega$, etc. have been found to simplify via the use of the Feynman scaling variable $x_F \equiv p_{\ell}/p_{\mathrm{max}}$ defined in the center of mass frame, where $p_{\ell}$ is the longitudinal momentum carried by the produced meson and $p_{\rm max}\approx\sqrt{s}/2$ at very large scattering energies. An alternative scaling variable $x_R \equiv E/E_{\rm max}$ in the center-of-mass frame has also been found to extend the range of validity of scaling at sub-asymptotic energies. 
Following \cite{BMPT:Bonesini:2001iz}, an efficient parametrization of the differential production cross-section may be presented as follows,
\begin{align}
E \frac{d^3\sigma}{dp^3}
& =\frac{E}{\pi p_{\rm max}}\frac{d^2\sigma}{dx_Fdp_T^2} \nnl
& \approx A \frac{(1-x_R)^\alpha}{x_R^\beta} (1+Cx_R)\exp(-Bp_T^2),
\label{BMPT}
\end{align}
where the exponents $\alpha$ and $\beta$ are fit to the data (conventionally presented in terms of $x_F$ variable~\cite{Aguilar-Benitez:1991hzq}). At all energies, the $p_T$ dependence out to $\sim 2 \, \mathrm{GeV}$ is well presented by the exponential form, where the measured value of the exponent is $B=2.6\pm 0.1 \, \mathrm{GeV}^{-2}$ for inclusive $\rho^0(775)$ meson at $400 \, \mathrm{GeV}$ beam energy~\cite{Aguilar-Benitez:1991hzq}.  Alternatively, the parametrization can be expressed by a similar variable $x_0\equiv p/p_{\rm max}=\sqrt{x_F^2+p_T^2/p_{\rm max}^2}$~\cite{Suzuki:1980fj}. 

\begin{figure}[t]
    \begin{center}
    \includegraphics[width=0.48\textwidth]{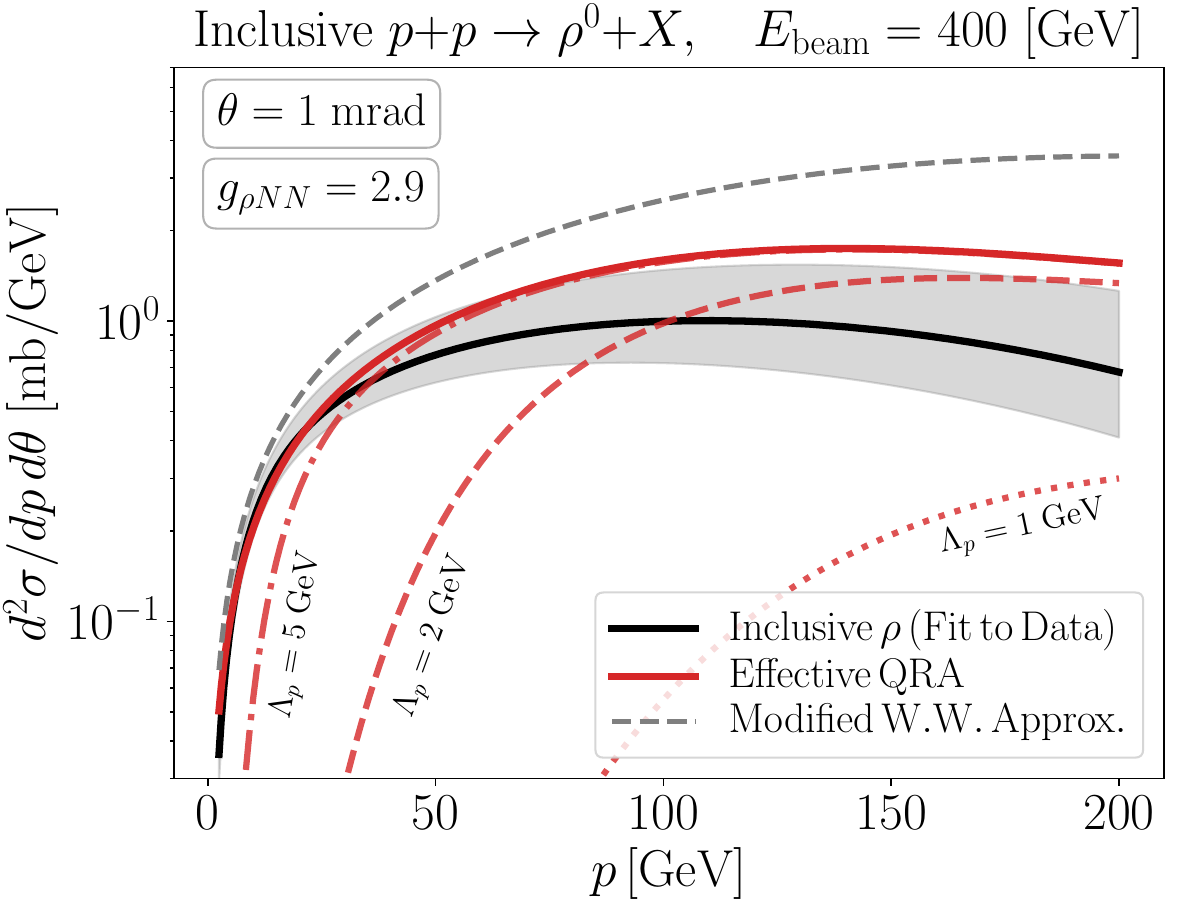}
    \caption{The momentum-angle distributions for inclusive $\rho (770)$ meson production, comparing data from the NA27 experiment~\cite{Aguilar-Benitez:1991hzq} (with a grey precision band) to the effective QRA and modified WW approaches. The choices of the $\Lambda_p$ scale indicate the impact of the off-shell form-factor in the effective QRA approach.}
    \label{fig:rho}
    \end{center}
\end{figure}

The best-fit parametrization for the differential cross-section in Eq.~(\ref{BMPT}) as a function of $x_F$ was determined by comparison with data for inclusive $\rho^0$ production from the NA27 experiment with $\sqrt{s} = 27.5 \, \text{GeV}$. The resulting parametrization was observed to lie within the error bars, and a confidence band was generated by varying the fit parameters within their uncertainties, resulting in a range of $d\sigma/dx_F$ curves consistent with the data. By changing the variables, the corresponding best-fit momentum-angle distribution as a function of momentum $p$ at a fixed angle is presented in~\cref{fig:rho}, in which the gray band illustrates the uncertainty in the fit.

Features of the distribution (\ref{BMPT}) have been motivated on the basis of Regge or string models for meson formation, reflecting the hadronic composite nature of the $\rho$. Thus, the large momentum asymptotics in particular, exhibiting a large suppression, are not expected to translate directly to the production of fundamental vectors such as dark photons. Nonetheless, a comparison for radiated momenta which are small relative to the beam momentum is informative as a test of models for proton bremsstrahlung and ISR in particular. The fit refers to inclusive $\rho$ production, and thus necessarily incorporates modes beyond ISR or bremsstrahlung more generally. Accordingly, it can viewed as providing an approximate upper bound on the rate obtained by QRA or other mechanisms, at least for low momenta. To evaluate the production of neutral $\rho$ mesons via the effective QRA approach, we use an effective $\rho$-nucleon vector coupling $\mathcal{L}_{\rm eff} = g_{\rho NN}\overline{N}\gamma^\mu \Vec{\rho}_\mu\cdot\Vec{\tau}N$ with an average experimental value of $g_{\rho NN}=2.9 \pm 0.3$~\cite{Oset:1983mvn,Downum:2006re,Riska:2000gd} and neglect the tensor coupling, having verified that its inclusion in the form $g_{\rho NN}\kappa_{\rho}/2m_N$ with $\kappa_{\rho}\sim 4$~\cite{Riska:2000gd} has negligible effect. A comparison of the various calculational approaches with the fit to data from NA27 is shown in \cref{fig:rho}. Considering the kinematic regime with small longitudinal and transverse momentum, we observe similar behaviour for the production rates from QRA, although the rate is suppressed by the off-shell form-factor, while the modified WW approximation starts to exceed the rho production data for larger momenta. Importantly, Fig.~\ref{fig:rho} suggests that choosing a $\Lambda_p$ scale around or slightly above the hadronic scale passes the test of not predicting an overproduction of $\rho$ mesons in the relevant kinematic range.

%%%%%%%%%
\subsection{Sensitivity to dark vectors}

\begin{figure}[t]
\centering
\includegraphics[width=0.48\textwidth]{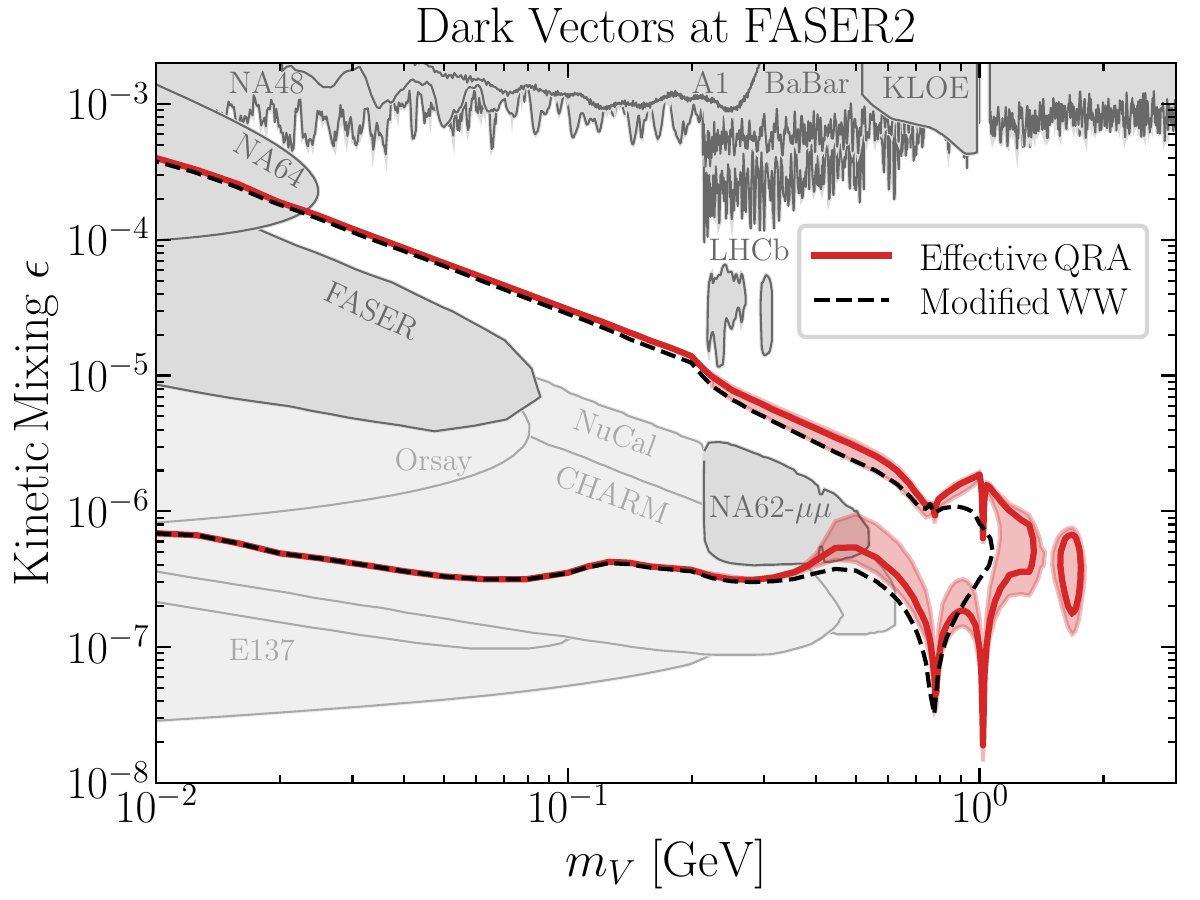}
    \caption{The sensitivity reaches for dark vectors decays at the FASER2 experiment, comparing bremsstrahlung production via the effective QRA (solid red) with the modified WW approximation (dashed black) with transverse momentum restricted to $p_T < 1 \gev$. This plot was created using the \texttt{FORESEE}~\cite{Kling:2021fwx} package. It incorporates various existing constraints along with the projections for FASER2 and considers additional dark vector production modes.}
    \label{fig:Sensitivity}
\end{figure}

\begin{figure}[t]
\centering
\includegraphics[width=0.48\textwidth]{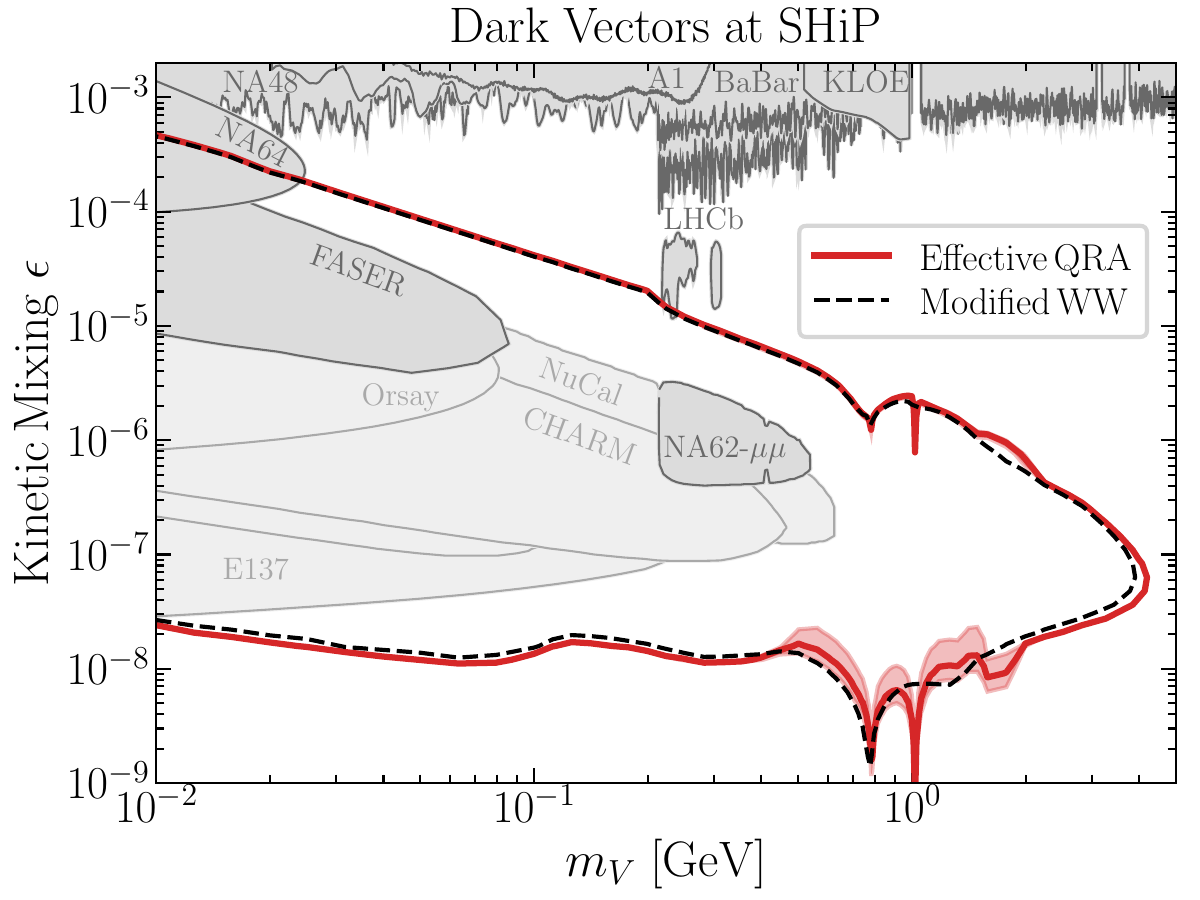}
    \caption{The sensitivity reaches for dark vectors decays at the SHiP experiment, comparing bremsstrahlung production via the effective QRA (solid red) with the modified WW approximation (dashed black), restricting the transverse momentum to $p_T < 4 \gev$ following~\cite{SHiP:2020vbd}. It incorporates various existing constraints along with the projections for SHiP and considers additional dark vector production modes including meson decays following the BMPT~\cite{BMPT:Bonesini:2001iz} production distribution and the Drell-Yan process at higher mass~\cite{SHiP:2020vbd}. }
    \label{fig:Sensitivity_SHiP}
\end{figure}

Finally, in this section, we illustrate the impact of the effective QRA production rate for proton bremsstrahlung on the sensitivity contours for the planned experiments SHiP and FASER2 in Figs.~\ref{fig:Sensitivity} and \ref{fig:Sensitivity_SHiP}. These experiments are used as representative examples of searches at high-energy fixed target experiments and the HL-LHC respectively. It is understood that there may also be impacts on the sensitvity at other existing or planned experiments (including NA62~\cite{NA62:2017rwk,Dobrich:2018ezn}, CHARM~\cite{CERN-Hamburg-Amsterdam-Rome-Moscow:1980ppk,Tsai:2019buq}, NuCal~\cite{Blumlein:1990ay,Tsai:2019buq}, DarkQuest~\cite{Apyan:2022tsd}, FACET~\cite{Cerci:2021nlb}, etc), but our goal is simply to illustrate the impact of the model of ISR production, and so we refrain from a more comprehensive analysis. 

In the following, we summarize some of the experimental details relevant to dark photon searches:
\begin{itemize}
    \item \textbf{SHiP}, the Search for Hidden Particles experiment at the ECN3 High-Intensity Beam Facility~\cite{Aberle:2839677}, plans to deliver $6 \times 10^{20}$ protons on a molybdenum alloy target as the hadron absorber. The experiment will utilize a muon shield that deflects muons from meson decay, creating an almost zero-background environment for searching for long-lived particles like dark photons. The Hidden Sector Decay Search (HSDS) detector, located about 33 meters from the target, features a 50 m-long vacuum decay vessel with a pyramidal frustum shape and a liquid scintillator veto system. The fiducial decay volume is designed to detect two charged tracks, reconstructing vertices from di-lepton and hadronic dark photon decays, followed by a spectrometer tracker and calorimeter. 
    
    \item \textbf{FASER2}, the ForwArd Search ExpeRiment at the High Luminosity LHC~\cite{Ariga:2018pin}, is a proposed experiment designed to search for light, weakly interacting particles such as dark photons. It will be situated within the Forward Physics Facility~\cite{Feng:2022inv}, approximately 620 meters downstream from the ATLAS interaction point, and will utilize the $14 \,\rm TeV$ LHC beam with $3 \,\rm ab^{-1}$ of integrated luminosity during the HL-LHC era. The FASER2 detector is compact, about 5 m long, and 1 m in diameter, and is equipped with high-resolution silicon strip detectors for precise tracking and a calorimeter system with tungsten or lead absorbers for energy measurement. This relatively inexpensive detector features two veto stations with scintillator layers that detect coincident muon tracks to minimize muon-induced backgrounds. It is anticipated that timing information will allow for the complete elimination of these backgrounds in the experiment.
\end{itemize}

We focus on the visible decay of dark photons into lepton pairs and heavier hadronic states, such as $\pi^+\pi^-$. The expected event rates of dark photon searches for achieving the projected sensitivity reaches for each experiment are computed given the differential cross sections of all channels as shown in Fig.~\ref{fig:Production}, convoluted with the survival/decay probability within the detector volume and considering the geometric acceptance of the detector. For SHiP, the vessel's acceptance probability ranges from $5 \%$ to $10 \%$, depending on the dark photon production mode~\cite{SHiP:2020vbd}. Reconstruction efficiency is near one for the bremsstrahlung channel but lower for meson decay channels due to the broader angular distribution of dark photons produced in these decays. A detailed discussion of these effects is beyond the scope of this paper and we refer the reader to Ref.~\cite{SHiP:2020vbd,FASER:2018eoc,FASER:2023tle} for more information. Figs.~\ref{fig:Sensitivity} and \ref{fig:Sensitivity_SHiP} indicate the resulting sensitivity contours compared to the earlier modified WW approach, making use of the \texttt{FORESEE}~\cite{Kling:2021fwx} package. Similar results can be obtained using \texttt{SensCalc}~\cite{Ovchynnikov:2023cry,KO24a}.

\section{Discussion}\label{sec:Discussion}
In this paper, we have refined the approach to modeling the forward production of dark vectors at proton colliders and fixed target experiments in the mass range from 0.5 to 2.0 GeV. Bremsstrahlung through ISR provides an important channel in this mass range, and we have improved the QRA approach to ISR via use of the Dawson correction to remove the unphysical $1/m_V^2$ singularity. This leads to a differential distribution (analogous to the one used for $q\rightarrow q'V$ for EW production in HERWIG) having kinematic features in common with the well-defined result for quasi-elastic radiation. We have also improved the analysis with a more complete treatment of the electromagnetic form factor at the proton vertex, including the dipole coupling, where resonant enhancements impact the sensitivity above the $\rho/\omega$ mass range. Fits to these form-factor contours are provided to enable straightforward implementation of this production model in other analyses \cite{fitfiles}. 

We have also attempted to benchmark the dark vector production model by computing the analogous QRA rate for $\rho$ meson production, and comparing it with inclusive data from NA27. While the comparison is not one-to-one as there are additional production modes, it indicates that the QRA rate lies below the inclusive data for low radiated momenta as required.

The uncertainty in the form factors was parametrized to provide some indication of the precision of the overall QRA rate, which is necessarily lower for vector masses above the well-measured $\rho/\omega/\phi$ resonances. It is natural to ask how this analysis might further be improved. For fixed target experiments, the scattering of protons off target neutrons is also significant, and could naturally be incorporated into the QRA analysis, along with (kinematically softer) secondary production modes. There is also the important question of how to incorporate FSR contributions, which for fully inelastic processes would require at least a statistical treatment of hadronization, or a full parton-level analysis. This may become feasible once PDFs are sufficiently well-constrained at small $x$.

\begin{acknowledgments}

We would like to express our thanks to F. Kling and M. Ovchynnikov for many helpful discussions and comments on the manuscript. We also thank M. Ovchynnikov for making us aware of the forthcoming analysis \cite{KO24a}. We are grateful to M. Hoferichter, A. Mart\'{i}nez Torres and K. P. Khemchandani for helpful correspondence and assistance with the calculation of the dispersive form factors, and also to S. Dubnicka for assistance with the UA model code. SF and AR acknowledge the support of NSERC, Canada, and PR is supported by Funda\c{c}\~ao de Amparo \`a Pesquisa do Estado de S\~ao Paulo (FAPESP) under the contract 2020/10004-7.

\end{acknowledgments}

\appendix

% change equation numbering
\counterwithin*{equation}{section}
\renewcommand\theequation{\thesection\arabic{equation}}
% changed equation numbering

\vspace{8mm}
{\centering \small \bf APPENDICES\\ }
%\vspace{5mm}

\section{Nucleon Electromagnetic Form Factors}
\label{app:EM_FF}

\begin{figure*}[t]
\centering
\includegraphics[width=0.49\textwidth]{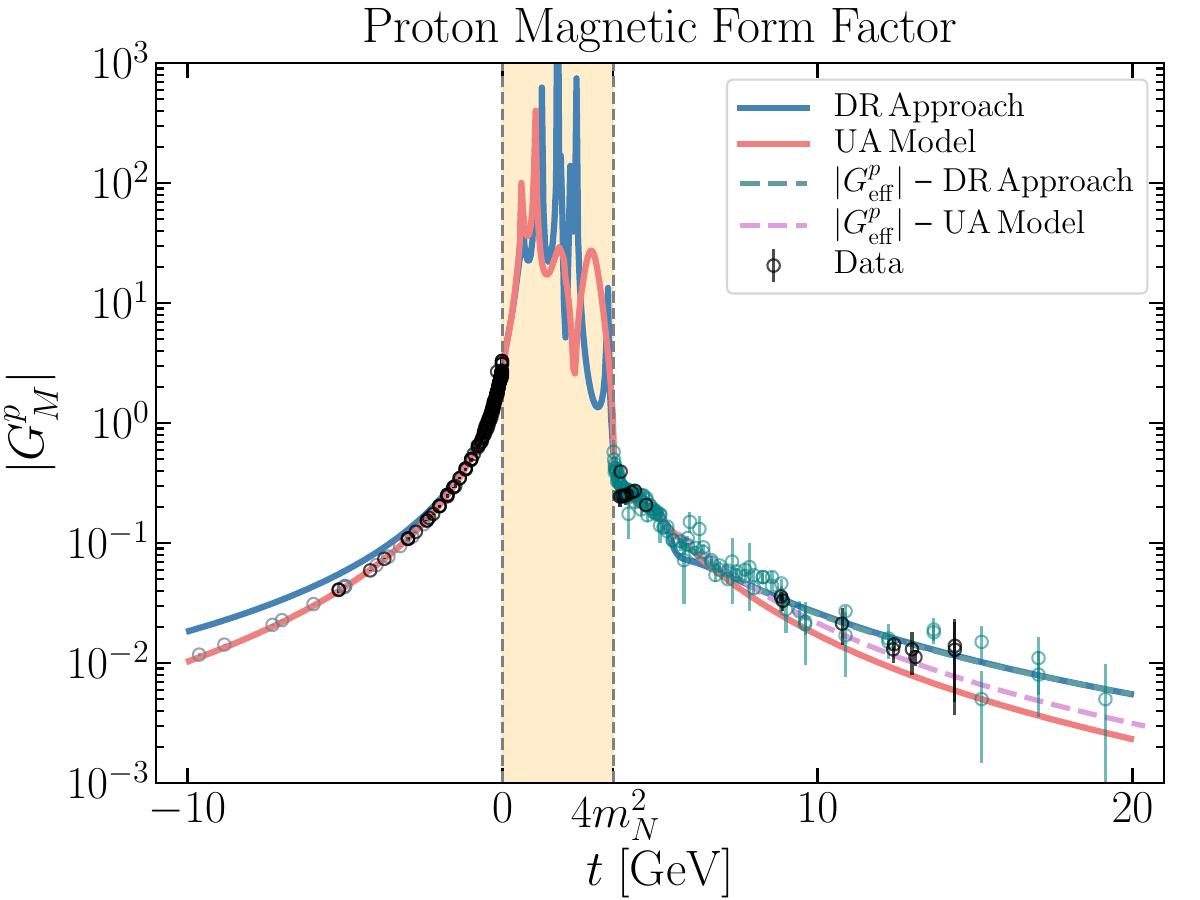} 
\includegraphics[width=0.49\textwidth]{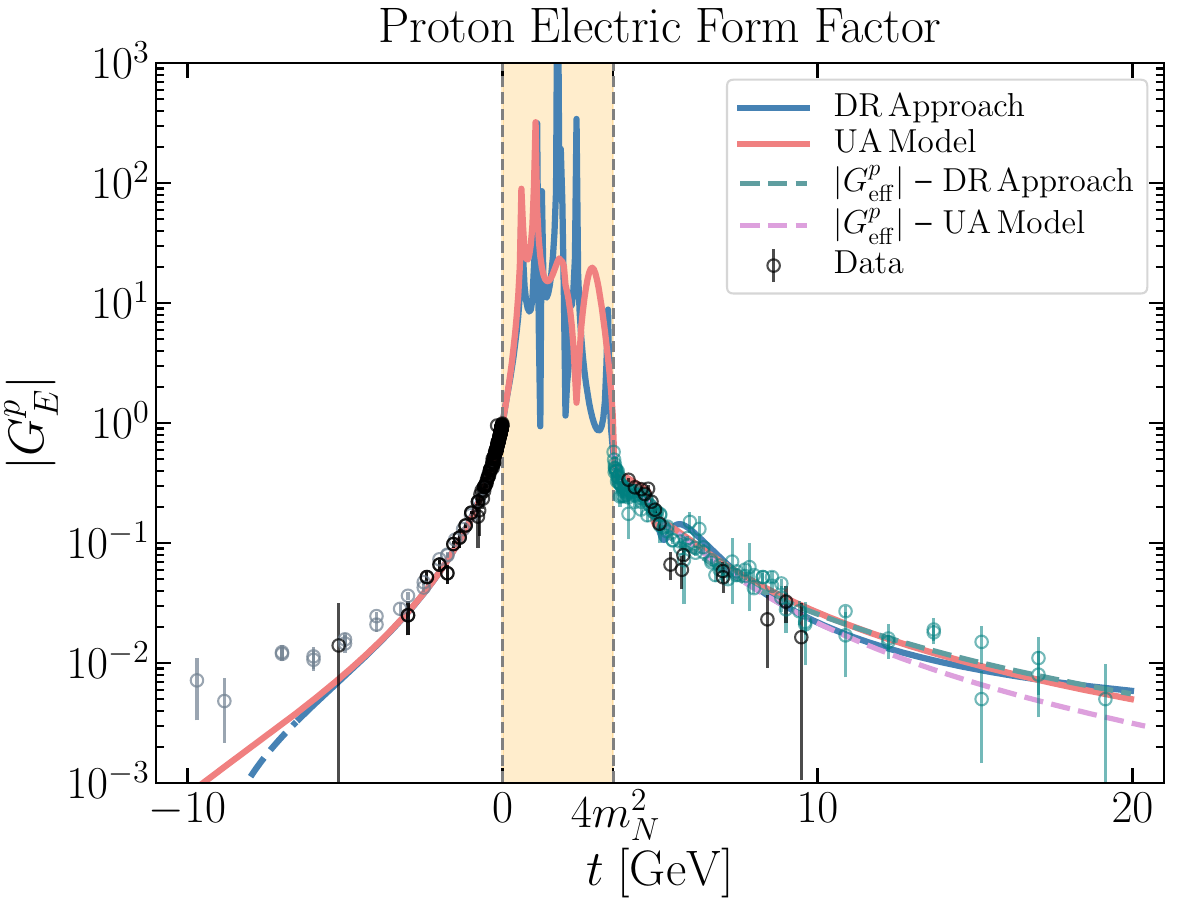} 
 \\
\vspace{0.1cm}
\includegraphics[width=0.49\textwidth]{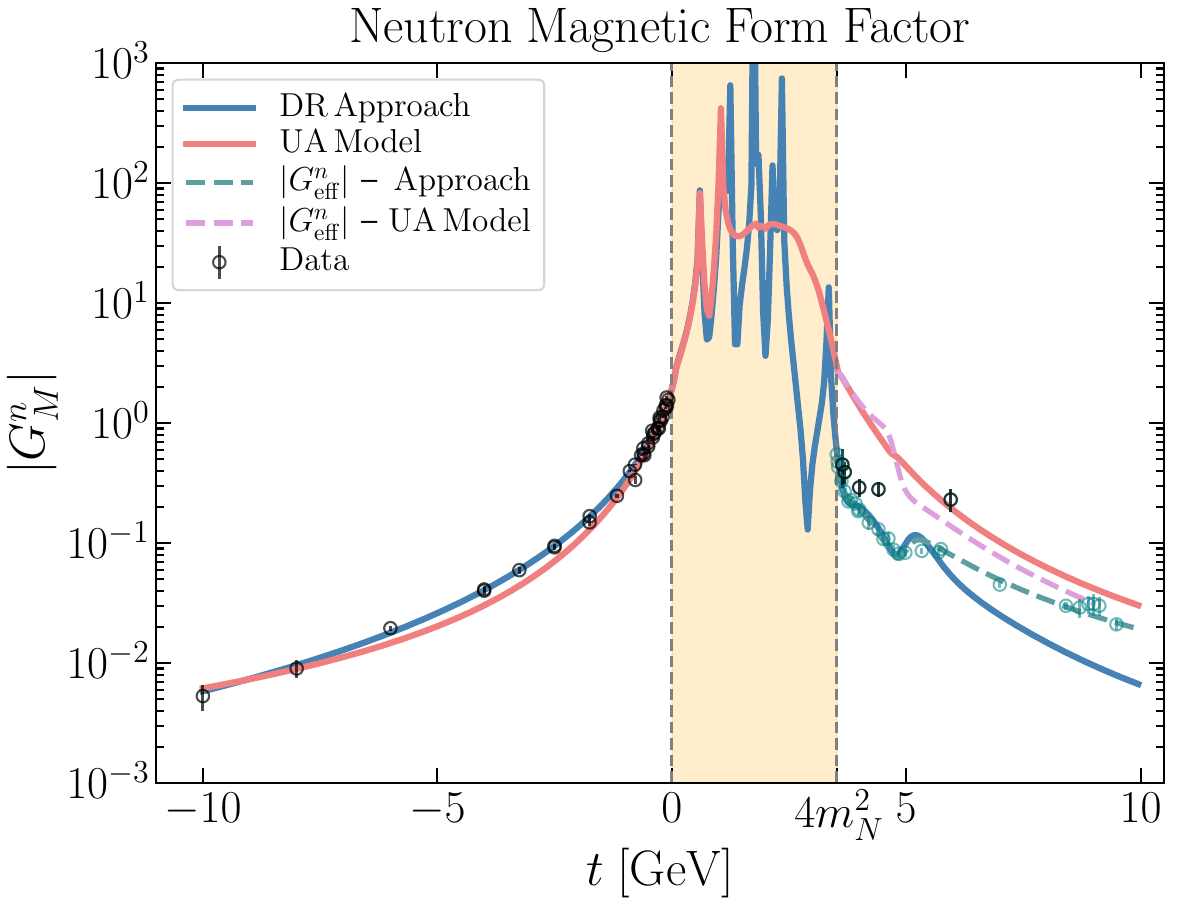}
\includegraphics[width=0.49\textwidth]{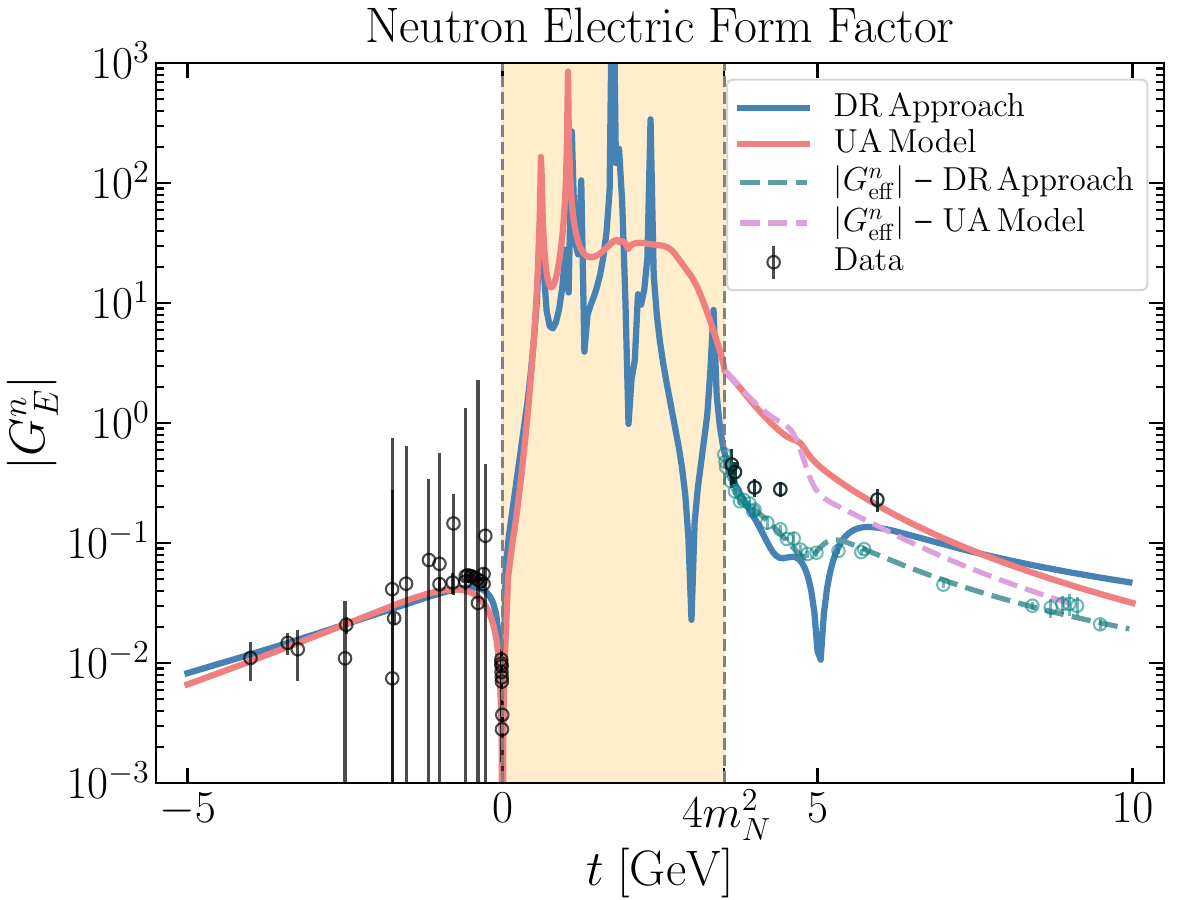} 
\caption{Fit to data for the electric and magnetic Sachs form-factors of the nucleon using both the dispersion relation analysis (blue), and unitary and analytic model (red) in both spacelike and timelike regions. In the timelike region, the dashed lines represent $G_{\rm eff}$ from the dispersion relation analysis (green) and the unitary and analytic model (pink). The spacelike data for the proton (gray) at high $Q^2$ are uncorrected and excluded from the fit. See the text for further details. }
\label{fig:EM_FF_fit}
\end{figure*}

The nucleon electromagnetic form factors can extracted from scattering data in the space-like region $(t<0)$, and from creation/annihilation processes in the physical timelike region ($t > 4m_N^2$). However, we require knowledge of the form-factor in the so-called `unphysical' timelike region $t \sim 1$ GeV$^2$ where data is missing, as indicated by the colored areas in Figs.~\ref{fig:EM_FF_fit}, which therefore requires a model and/or careful spectral analyses. We will consider two approaches in this appendix.

In both approaches, the nucleon form factor is normalized 
\begin{align}
& F_1^p(0)=1, \quad F_1^n(0)=0 \\
& F_2^p(0)=\mu_p-1, \quad F_2^n(0)=\mu_n
\end{align}
with $\mu_p = 2.793$ and $\mu_n = - 1.913$ the magnetic
moment of the proton and the neutron, respectively. At large momentum transfer, the asymptotic form of the Dirac and Pauli form factors is prescribed by perturbative QCD~\cite{Brodsky:1974vy,Lepage:1980fj}, 
\begin{equation}
\lim _{t \rightarrow \infty} F_i\left(t\right) \sim \frac{1}{t^{i+1}}, \quad i=1,2,
\end{equation}
where logarithmic corrections~\cite{Belitsky:2002kj} are neglected.

Experimental data is often presented in terms of the electric and magnetic Sachs form factors $G_E$ and $G_M$ which are related to $F_{1,2}$ as follows (with $N$ labeling the nucleon),
\begin{align}
    G^N_E(t) &\equiv F^N_1(t) + \frac{t}{4m_N^2} F^N_2(t), \\
    G^N_M(t) &\equiv F^N_1(t) + F^N_2(t).
\end{align}

The timelike data is also presented in terms of the effective form factor
\begin{equation}
|G^N_{\text{eff}}| = \sqrt{\frac{|G^N_E|^2 + \xi |G^N_M|^2}{1 + \xi}},
\end{equation}
with $\xi = t/2m_N^2$ (see Ref.~\cite{Lin:2021xrc}). In the space-like region, a successful method of determining the Sachs form factors has been the measurement of recoil polarizations that allows one to measure the ratio
\begin{align}
    R_p\equiv \mu_p \frac{G_E^p}{G_M^p}~.
\end{align}
In this ratio, many systematic uncertainties cancel out, making this observable more robust than the Sachs form-factors themselves, as indicated by gray data points in Fig.~\ref{fig:EM_FF_fit}.

\subsection{Unitary-Analytic Model}
Unitarity implies that nucleon electromagnetic form factors are analytic functions throughout the entire complex $t$-plane, except for cuts extending along the positive real axis from the lowest continuum branch point $t_0$ to infinity. This property has been used to determine a unitary and analytic (UA) template for the form factors with parameters determined via scattering data and knowledge of the vector meson resonances. This UA model unifies the vector-meson pole contributions and effective cut structures~\cite{Adamuscin:2016rer}. The true neutral vector-mesons, including $\rho(770)$, $\omega(782)$, $\phi(1020)$, $\rho(1450)$, $\omega(1420)$, $\phi(1680)$, $\rho(1700)$, $\omega(1650)$, and $\phi(2170)$ \cite{PDG2016} are combined in a series of poles within the framework of the Vector Meson Dominance (VMD) model, in the generic form
\be
 F(t) = \sum_{r=1}^n \frac{f_{V_rNN}}{f_{V_r}}\frac{m_{V_r}^2}{m_{V_r}^2-t},
\label{eq:UAFF}
\ee
with the poles broadened e.g. via the replacement $m_{V_r}^2 \rightarrow m_{V_r}^2-i\Gamma_{V_r}/2$.
The cut structures are implemented via a conformal mapping, which transforms the cut starting at $t_{\text{cut}}=4(9)m_{\pi}^2$ for the iso-vector (iso-scalar) case in the $t$-plane onto the unit circle using a new variable
\begin{equation}
V\left(t\right)=\frac{\sqrt{t_{\text {in }}{-}t_{\text {cut}}}-\sqrt{t_{\text {in }}{-}t}}{\sqrt{t_{\text {cut}}{-}t}},
\end{equation}
with the corresponding inverse transformation taking the nonlinear form
$t=t_{\text {cut }}+4\left(t_{\text {in }}{-}t_{\text {cut }}\right) V^2(t)/\big(V^2(t){-}1\big)^2$, where $t_{\rm in}$ parametrize effective inelastic thresholds.
This non-linear transformation for $t$ implies the relations $m_r^2=t_{\rm cut}+\frac{4(t_{\rm in}-t_{\rm cut})}{[1/V_r-V_r]^2}$ and $0=t_{\rm cut}+\frac{4(t_{\rm in}-t_{\rm cut})}{[1/V_N-V_N]^2}$, where $V_N = V(0)$ and $V_r=V(m_r^2)$. Applying these relations to Eq.~(\ref{eq:UAFF}), the pole terms take the form
\begin{equation}
    \frac{m_r^2-0}{m_r^2-t} \to \big(\frac{1{-}V^2}{1{-}V_N^2}\big)^2\bigg( 
\frac{(V_N^2{-}V_r^2)(V_N^2{+}1/V_r^2)}{(V^2{-}V_r^2)(V^2{+}1/V_r^2)} \bigg).
\end{equation}

Depending on the pole location relative to the effective inelastic threshold $t_{\rm in}$, the introduction of a nonzero resonance width $m_r^2 \rightarrow (m_r^2-i\Gamma_r/2)^2$ modifies the pole expression to $(1{-}V^2)^2/(1{-}V_N^2)^2 L_r(V)$, with $L_r$ a complex function given in \cite{Adamuscin:2016rer}.

For the isoscalar and isovector Dirac and Pauli form factors, the values of $t_{\rm in}$ and the coefficients $\frac{f_{V_rNN}}{f_{V_r}}$ in the series (\ref{eq:UAFF}) are free parameters determined by comparing the model with existing data in spacelike and timelike regions simultaneously.

\subsection{Dispersion Relation Analysis}
In the Dispersion Relation (DR) approach~\cite{Lin:2021xrc,Lin:2021umz,Belushkin:2006qa}, the spectral function $\mathrm{Im} \, F(t)$ serves as the basic parameterization of physical effects contributing to the nucleon form factors. The general form of the spectral function permitted by unitarity consists of a combination of continua and poles. The low-mass continua encompass $2\pi$, $K\bar{K}$, and $\rho \pi$ contributions. Effective vector meson poles are used to approximate higher mass continua. 

The complete isoscalar and isovector parts of the Dirac and Pauli form factors can then be parameterized as follows~\cite{Hoferichter:2016duk,Lin:2021umz},
\begin{align}\label{eq:FormFactorscalarvector}
F_i^s(t) & =F_i^{(s,K\bar{K})}(t) + F_i^{(s,\rho\pi)}(t) + \sum_{V=\omega,\phi,s_1, . .} \frac{a_i^V}{m_V^2-t}, \nonumber \\ 
F_i^v(t) & =F_i^{(v,\pi\pi)}(t) + \sum_{V=v_1, v_2, . .} \frac{a_i^V}{m_V^2-t},
\end{align}
where $i=1,2$, and with poles again broadened e.g. via the replacement $m_{V_r}^2 \rightarrow m_{V_r}^2-i\Gamma_{V_r}/2$. The low-mass pole terms in (\ref{eq:FormFactorscalarvector}) correspond to physical vector mesons, namely the $\omega(782)$ and the $\phi(1020)$, whereas the higher mass poles ($s_1,s_2,...$ for isoscalar and $v_1,v_2,...$ for isovector channels) serve as effective parameters to account for unknown continuum contributions.  The two-pion exchange continuum in the isovector form factor is derived using the pion form factor and the $\pi\pi\rightarrow N\bar{N}$ partial waves, which naturally exhibit the $\rho$-resonance along with a notable enhancement on the left shoulder of the resonance. 
We use the ancillary files of Ref.~\cite{Hoferichter:2016duk} for the isovector spectral functions which include results from $\pi\pi\rightarrow N\bar{N}$ partial waves extracted from Roy-Steiner equations, consistent input for the pion vector form factor $F_\pi^V(t)$, and a discussion of isospin-violating effects.

The lowest isoscalar continuum, given by the 3-pion exchange, has been shown in chiral perturbation theory to be negligible such that there is no enhancement on the left wing of the $\omega$ resonance~\cite{Bernard:1996cc}. The next most important contributions are from $K\bar{K}$ and $\rho \pi$ continua that can be represented by an effective pole term at the $\phi(1020)$~\cite{Belushkin:2006qa} and fictitious $\omega^\prime$ meson with a mass $M_{\omega^\prime} = 1.12 \gev$~\cite{Meissner:1997qt}, respectively. Additional continua arising from a higher number of pions experience strong suppression. Given that the fit was already performed with the most up-to-date data sets, both in the space- and time-like region, even including neutron data, we refrain from performing our analysis and use the fit results of Ref.~\cite{Lin:2021xrc}.

The results of both UA and DR approaches are compared to data in Fig.~\ref{fig:EM_FF_fit}. While the fits are quite good in the spacelike and physical timelike domains, the two approaches diverge in the unphysical shaded region. We find that the UA model, constrained by the map to observed resonances, produces form factors that are somewhat lower overall, and so to be conservative we adopt this model in the rate analysis carried out in the paper.

\bibliography{References}
\end{document}